\documentclass[fleqn,usenatbib]{mnras}

\usepackage{newtxtext,newtxmath}

\usepackage[T1]{fontenc}
\usepackage{ae,aecompl}

\usepackage{graphicx}	
\usepackage{amsmath}	

\usepackage{enumitem}
\usepackage{bm}

\usepackage{etoolbox}
\usepackage[dvipsnames]{xcolor}

\usepackage{CJKutf8}
\newcommand{\chn}[1]{\begin{CJK}{UTF8}{gbsn}#1\end{CJK}}

\makeatletter
\patchcmd\@combinedblfloats{\box\@outputbox}{\unvbox\@outputbox}{}{%
}%
\makeatother

\makeatletter
\newcommand*{\rom}[1]{\expandafter\@slowromancap\romannumeral #1@}
\makeatother

\newcommand{\orcid}[1]{\hskip1pt \textsuperscript{\href{https://orcid.org/#1}{\includegraphics[width=7pt]{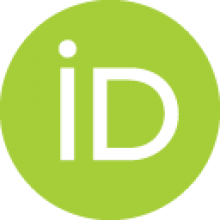}}}}

\newcommand\mystack[2]{\genfrac{}{}{0pt}{}{#1}{#2}}

\title[Torus-stable zone]{Torus-stable zone above starspots}

\author[X. Sun et al.]{
Xudong Sun (\chn{孙旭东})\orcid{0000-0003-4043-616X},$^{1}$\thanks{E-mail: xudongs@hawaii.edu (XS)}
Tibor T\"{o}r\"{o}k\orcid{0000-0003-3843-3242},$^{2}$
and Marc L. DeRosa\orcid{0000-0002-6338-0691}$^{3}$
\vspace{2mm}\\
$^{1}$Institute for Astronomy, University of Hawai`i at M\={a}noa, Pukalani, HI 96768, USA\\
$^{2}$Predictive Science Inc., San Diego, CA 92121, USA\\
$^{3}$Lockheed Martin Solar and Astrophysics Laboratory, Palo Alto, CA 94304, USA
}

\date{Accepted 2021 November 5. Received 2021 November 4; in original form 2021 September 14}

\pubyear{2021}

\begin{document}
\label{firstpage}
\pagerange{\pageref{firstpage}--\pageref{lastpage}}
\maketitle

\begin{abstract}
Whilst intense solar flares are almost always accompanied by a coronal mass ejection (CME), reports on stellar CMEs are rare, despite the frequent detection of stellar `super flares'. The torus instability of magnetic flux ropes is believed to be one of the main driving mechanisms of solar CMEs. Suppression of the torus instability, due to a confining background coronal magnetic field that decreases sufficiently slowly with height, may contribute to the lack of stellar CME detection. Here we use the solar magnetic field as a template to estimate the vertical extent of this `torus-stable zone' (TSZ) above a stellar active region. For an idealised potential field model comprising the fields of a local bipole (mimicking a pair of starspots) and a global dipole, we show that the upper bound of the TSZ increases with the bipole size, the dipole strength, and the source surface radius where the coronal field becomes radial. The boundaries of the TSZ depend on the interplay between the spots' and the dipole's magnetic fields, which provide the local- and global-scale confinement, respectively. They range from about half the bipole size to a significant fraction of the stellar radius. For smaller spots and an intermediate dipole field, a secondary TSZ arises at a higher altitude, which may increase the likelihood of `failed eruptions'. Our results suggest that the low apparent CME occurrence rate on cool stars is, at least partially, due to the presence of extended TSZs.
\end{abstract}




\section{Introduction}
\label{sec:intro}


\subsection{Torus instability in solar eruptions}
\label{subsec:intro_ti}

Solar coronal mass ejections (CMEs) are rapid expulsions of magnetized plasma with velocity up to $3000$\,km\,s$^{-1}$ and mass up to a few $10^{16}$\,g \citep{webb2012}. The fastest CMEs are often associated with intense flaring  \citep{vrsnak2005}. In these events, the CME kinetic energy and the flare radiative energy are generally both of order $10^{32}$--$10^{33}$~erg \citep{emslie2012}. The interaction between a CME and the Earth's magnetosphere can cause severe space weather disturbances \citep{baker2016}.

Many solar CMEs originate from active regions that harbour kilogauss magnetic fields and sunspots. Prior to eruption, their coronal fields are thought to often evolve towards a `magnetic flux rope' configuration, i.e. a current-carrying magnetic flux tube with coherent twist \citep{patsourakos2020}. Evidence of pre-eruption flux ropes in the low corona has been reported frequently \citep[e.g.][]{green2011,patsourakos2013,duan2019,kliem2021}. The twisted structure is clearly visible in a significant fraction of CMEs \citep{vourlidas2013}.

\begin{figure}
\centerline{\includegraphics{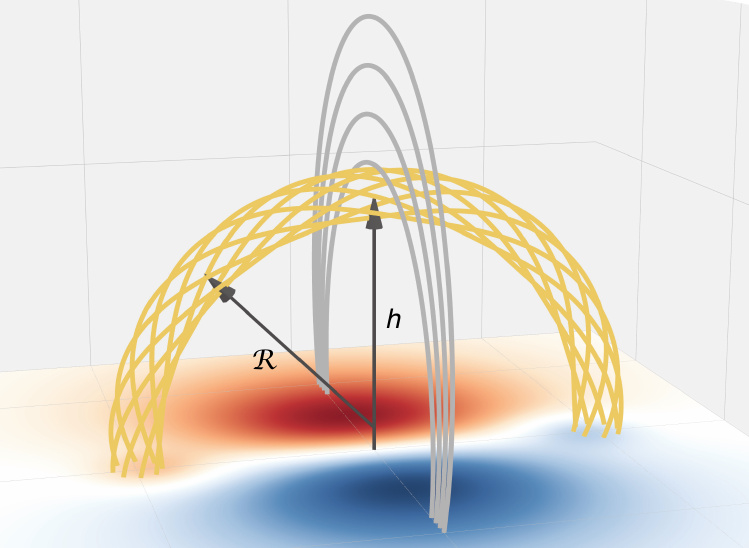}}
\caption{Illustration of the basic magnetic ingredients for the torus instability. The yellow field lines form a partial toroidal flux rope, whose toroidal current is clockwise directed. The grey field lines indicate the strapping field. The red (blue) colour shows positive (negative) photospheric vertical field. The height of the apex $h$ and the major radius of the torus $\mathcal{R}$ are marked.}
\label{f:torus}
\end{figure}

The torus instability \citep{bateman1978,kliem2006} is believed to be one of the main driving mechanisms of solar CMEs. For a toroidal flux rope \citep[Fig.~\ref{f:torus}; see also Fig.~1 of][]{myers2016}, the internal poloidal magnetic field and toroidal electric current produce an outward-directed force, known as the `hoop force', that drives its expxansion. The interaction of the flux rope's current with an external poloidal field $\bm{B}_p$, also known as the `strapping field', generates an oppositely directed `strapping force' that provides the equilibrium and can stabilize the system. The torus instability sets in when $B_p=\lVert \bm{B}_p \rVert$ decreases fast enough with respect to the major radius $\mathcal{R}$ of the toroidal flux rope, so that it can no longer provide sufficient strapping force. For $B_p \propto \mathcal{R}^{-n}$, the `decay index' $n$ should satisfy
\begin{equation}
n = -\dfrac{\partial \ln B_p}{\partial \ln \mathcal{R}} > n_c,
\label{eq:n}
\end{equation}
where $n_c$ is a critical decay index, above which the torus instability may occur. Conversely, the torus instability will be suppressed if $n<n_c$. On the Sun, the coronal flux ropes are akin to partial tori whose legs are anchored in the photosphere \citep{chenj2003}. The height of the flux rope axis' apex above the photosphere, $h$, is commonly used in place of $\mathcal{R}$.

The corona may be divided into `torus-stable' ($n<n_c$) and `torus-unstable' zones ($n>n_c$), separated by a critical height $h_c$ (at which $n=n_c$). The value of $n_c$ depends on the details of the system, such as the aspect ratio of the flux rope, the profile of the toroidal current, etc. A range of $n_c$ between $0.5$ and $2$ has been found for solar cases based on analytical models \citep{isenberg2007,olmedo2010}, magnetohydrodynamic (MHD) simulations \citep{torok2007,aulanier2010,fan2010,kliem2014,zuccarello2015}, and laboratory plasma experiments \citep{myers2015,alt2021}. In particular, the value $n_c=1.5$, derived for a thin, axisymmetric, full torus \citep{bateman1978}, has been reproduced in several aforementioned MHD simulations, and is widely used in observational studies.

The majority of solar CMEs exhibit a slow expansion followed by an impulsive acceleration \citep{zhang2001,vrsnak2001}. For the torus instability to be triggered, the flux rope must first ascend from a torus-stable zone (TSZ) into an unstable one. This initial slow rise of the ejecta may be due to a number of mechanisms, such as flux cancellation, shearing or twisting motions \citep{green2018}, that facilitate (slow) tether-cutting \citep{moore2001} or breakout \citep{antiochos1999} reconnection. Indeed, the values of $n$ at the eruption onset height based on coronal field models are broadly consistent with the theoretical $n_c$ values \citep[e.g.][]{filippov2001,mccauley2015}. The occurrence of the torus instability is particularly supported when the signatures of alternative mechanisms, for instance flare emission due to magnetic reconnection, lag behind the CME acceleration or are largely absent \citep[e.g.][]{song2013,cheng2020}.

Observations show that some upward-moving flux ropes decelerate and come to a halt with no CME ensuing \citep{ji2003,green2007,chenhd2013,zhouzj2019}. Such a `failed eruption' may suggest that the eruption is initiated in a TSZ due to alternative mechanisms, such as the helical kink instability \citep{torok2005,hassanin2016} or `flare reconnection' \citep{karpen2012}. In these cases, the early kinematic evolution resembles that of a successful eruption; the slow-decaying background field is believed to play an important role in suppressing the eruption \citep{huangzw2020}. To test this hypothesis, many studies have compared modelled coronal fields prior to flares that occurred with or without a CME. The models for the latter type of flares, of which many are associated with failed eruptions, tend to have a stronger background field or a smaller $n$ at typical eruption onset heights \citep{wangy2007,liuy2008,cheng2011}, a smaller $n$ at the flux rope axis' apex \citep{jing2018,duan2019}, or a greater $h_c$ for a given $n_c$ \citep{wangd2017,sarkar2018,baumgartner2018,vasantharaju2018,lit2020}.

A failed eruption may also occur if a TSZ is situated above a torus-unstable zone. A flux rope may erupt due to the torus instability in the lower layer, and subsequently get trapped in the higher one. Indeed, some multipolar solar active regions exhibit a saddle-like $n(h)$ profile that crosses $n=n_c$ three times \citep{guo2010,cheng2011,wangd2017,filippov2020}. The three critical heights divide the corona into four layers; a secondary TSZ arises at a higher altitude.


\subsection{The `missing' stellar CMEs}
\label{subsec:intro_stcme}

Stellar flares are common on cool stars, and are thought to be powered by excess magnetic energy as on the Sun \citep{benz2010}. The energy-frequency relation follows a similar power law as the solar flares, but extends to much higher energy ranges \citep{shibata2013}. Thousands of `super flares' with $10^{33}$--$10^{36}$~erg bolometric energy have been detected, many of which occurred on Sun-like stars \citep[e.g.][]{schaefer2000,maehara2012,davenport2016,howard2019,gunther2020}.

Yet, whilst stellar flares are observed with regularity, stellar CME observations are rare. Less than $40$ candidates have been identified using the signatures of Doppler shift, X-ray absorption, or coronal dimming in extreme ultraviolet and X-ray \citep[e.g.][]{moschou2019,argiroffi2019,veronig2021}. Surveys of Balmer-line spectra find no signatures of CMEs in a sample of F--K dwarfs \citep{leitzinger2020}, and only weak line asymmetries on M dwarfs where the inferred velocities are mostly below the escape threshold \citep{vida2019,muheki2020}. Recent radio-optical observing campaigns report only one case of a type-\rom{4} radio burst \citep{zic2020} and no type-\rom{2} radio bursts \citep{crosley2018a,crosley2018b,villadsen2019}, the latter being a hallmark for CME-generated shocks on the Sun.

The lack of CME detection from cool stars is surprising. For the Sun, the flare-CME association rate $P$ increases with the soft X-ray flux $F_{\mathrm{SXR}}$ (as measured by the \textit{GOES} satellite's $0.1$--$0.8$~nm band), and reaches unity for the largest events where $F_{\mathrm{SXR}} \ge 10^{-3.5}$\,W\,m$^{-2}$ \citep[bolometric radiated energy $E$$\,\sim\,$$10^{32}$~erg;][]{andrews2003,yashiro2006}. Barring major observational biases, the stellar CME rate or their typical velocities must be significantly lower than the corresponding solar values. This will have major implications on the stellar mass loss and angular momentum loss rates \citep{drake2013,odert2017,wood2021} as well as on exoplanet habitability \citep{khodachenko2007,airapetian2020}.


\subsection{Magnetic confinement on cool stars}
\label{subsec:intro_confine}

NOAA solar active region 12192 in October 2014 hosted the largest sunspot group since 1990. It was the second most flare-productive region of solar cycle 24. Surprisingly, none of its major flares were associated with a CME. Analyses revealed a weaker magnetic twist in the core-region field and a stronger background field than that of two other flare-CME-rich regions \citep{sun2015}.

Active region 12192 serves as a solar analogue to the `missing stellar CME conundrum', to which failed eruptions may be a solution \citep{drake2016,osten2017}. Therein, magnetic reconnection driven by the initial eruption onset could explain the flaring; confinement of the eruption by the overlying field could explain the lack of CME detection.

Whilst the task of observing stellar flux ropes remains intractable, the large-scale stellar photospheric fields have been probed for a sizable sample of cool stars via spectropolarimetry observations \citep[e.g.][]{marsden2014,petit2014,folsom2016}. Inversion techniques such as the Zeeman Doppler Imaging \citep{semel1989,donati1997} and the Magnetic Doppler Imaging \citep{piskunov2002} have been used to infer low-resolution stellar magnetic field maps. The field strength and topology may be very different from those of the Sun, and potentially more effective at suppressing the CMEs. Here, we posit three relevant factors.

First, many cool stars host large spots over ten degrees in diameter with kilogauss magnetic fields, compared to typical sunspots of sometimes similar strength but diameters of just a few degrees \citep{berdyugina2005,strassmeier2009}. The unsigned magnetic flux $\lvert \Phi \rvert$ of such starspots will be much greater than the typical solar value, $10^{22}$~Mx, and can, therefore, provide strong `local confinement' in the low corona. This is supported by recent surveys of solar flares, where the flare-CME association rate $P$ is found to be anti-correlated with $\log \,\lvert \Phi \rvert$ \citep{lit2020}. For the empirical relation $P=\alpha \log F_{\mathrm{SXR}} + \beta$, the positive slope $\alpha$ is found to drastically decrease with $\lvert \Phi \rvert$ \citep{lit2021}. If this trend holds, even superflares with $F_{\mathrm{SXR}} = 0.01$\,W\,m$^{-2}$ ($E$$\,\sim\,$$10^{34}$~erg) are expected to have $P<0.5$ in a large stellar active region, $\lvert \Phi \rvert = 10^{24}$\,Mx. 

Second, some cool stars can have strong global-scale magnetic fields up to the kilogauss range, compared to just a few gauss on the Sun \citep{donati2009,reiners2012,kochukhov2021}. The efficacy of this `global confinement' is demonstrated with MHD modelling of stellar coronae based on a solar template. In \citet{alvaradogomez2018}, it is shown that a $75$~G dipole field is able to fully entrap CMEs with up to $3\times10^{32}$~erg kinetic energy. It is interesting to note that these failed eruptions are still accompanied by intense magnetic reconnection, which will produce flare radiative energy of the same order.

Third, it has been argued that cool stars with stronger surface magnetic activity will have a smaller fraction of magnetic flux open to the asterosphere \citep{schrijver2003,farrish2019}. Consequently, the larger fraction of closed magnetic flux is expected to exert stronger confinement on erupting flux ropes. In the widely used potential field source surface (PFSS) model for solar and stellar coronal fields \citep{altschuler1969,schatten1969}, the effect is approximated by higher `source surface' radii where the fields become open and radial due to the action of stellar winds \citep{reville2015,reville2016,see2017,see2018}.


\subsection{Aim and outline}
\label{subsec:intro_outline}

In the study presented here, we estimate the upper bound of the TSZ, i.e. the critical height $h_c$, above bipolar magnetic spots on cool stars. An extended TSZ is expected to reduce the likelihood of the torus instability onset and to increase the likelihood of failed eruptions. The suppression of the torus instability may help to explain the low apparent stellar CME rate.

To this end, we consider an idealised stellar magnetic environment comprising a local bipole (mimicking a pair of starspots) and a global dipole, which provide the local- and global-scale confinement, respectively. Henceforth, the starspot field configuration will be referred to as a (local) `bipole', whereas the background stellar field configuration will be referred to as a (global) `dipole'. We use the PFSS model to calculate the large-scale coronal magnetic field that an erupting flux rope would have to encounter, and investigate how the TSZ above starspots varies with the bipole size, the dipole strength, and the source surface radius.

The remainder of the paper is organized as follows. In Section~\ref{sec:method}, we describe the idealised model, and present the results of the model calculations in Section~\ref{sec:result}, followed by a discussion of the limitations of the model and the implications of the results in Section~\ref{sec:disc}.


\section{Method}
\label{sec:method}


\subsection{PFSS model}
\label{subsec:pfss}

We adopt an asterographic coordinate system specified by radius $r$, colatitude $\theta\in[0,\pi]$, and longitude $\phi\in[-\pi,\pi]$. The PFSS model solves for the coronal magnetic field $\bm{B}(r,\theta,\phi)$ between the stellar surface $r=R_\star$ and the source surface $r=R_s$. The field is divergence- free (from Maxwell's equations) and is assumed in the PFSS model to also be current-free and thus curl-free. As a result, it can be described by a scalar potential $\Psi(r,\theta,\phi)$ that satisfies
\begin{equation}
\begin{split}
\nabla\Psi &= -\bm{B}, \\
\nabla^2 \Psi &= 0. \\
\end{split}
\label{eq:psi_def}
\end{equation}
The inner boundary condition at $r=R_\star$ is commonly set to match the photospheric radial field \citep{wang1992}. Additionally, $\Psi$ must be constant (usually zero) on the outer boundary at $r=R_s$ as the field is assumed to be radial there. That is,
\begin{equation}
\begin{split}
\left. \dfrac{\partial \Psi}{\partial r}\right\vert_{r=R_\star}  &=  -B_r(R_\star,\theta,\phi), \\
\left. \Psi \right\vert_{r=R_s}  &=  0. \\
\end{split}
\label{eq:bd}
\end{equation}

The solution can be written as \citep{hoeksema1984}:
\begin{equation}
\begin{split}
\Psi = R_\star & \sum_{\ell=0}^{\infty} \sum_{m=0}^{\ell} P_\ell^m(\cos\theta) \, \left( g_{\ell m} \cos {m\phi} + h_{\ell m} \sin {m\phi} \right) \times \\
& {\left(\dfrac{R_\star}{r}\right)^{\ell+1} \left[ 1 - \left( \dfrac{r}{R_s} \right)^{2\ell+1} \right]} \left/ {\left[ \ell +1+ \ell \left(\dfrac{R_\star}{R_s}\right)^{2\ell+1} \right]} \right., \\
\end{split}
\label{eq:psi}
\end{equation}
where $\ell$ and $m$ are respectively the degree and order of the spherical harmonics expressed in part as the associated Legendre polynomials, $P_\ell^m(x)$. The coefficients $g_{\ell m}$ and $h_{\ell m}$ can be determined by integrating the surface field:
\begin{equation}
\mystack{g_{\ell m}}{h_{\ell m}} = c_\ell \int\limits^{\pi}_{-\pi} \mystack{\cos}{\sin}m\phi\,\mathrm{d}\phi \int\limits^{\pi}_0 B_r(R_\star,\theta,\phi)\,P^m_\ell(\cos\theta)\sin\theta\,\mathrm{d}\theta,
\label{eq:ghint}
\end{equation}
where the multiplier $c_\ell=(2\ell+1)/4\pi$ is a result from the Schmidt quasi-normalization scheme for the spherical harmonics. An axial dipole field is represented by the ($\ell=1$, $m=0$) mode, for which $B_r$ at the stellar surface is $g_{10}\cos\theta$, with the amplitude of this mode  proportional to $g_{10}$.

We can evaluate the energy spectrum of the surface $B_r$ using the mean squared field associated with a certain $\ell$ \citep{derosa2012}:
\begin{equation}
\langle B^2_{r,\ell} \rangle = \dfrac{1}{2\ell+1} \sum_{m=0}^\ell \left( g^2_{\ell m} + h^2_{\ell m} \right).
\label{eq:bl2}
\end{equation}


\subsection{Bipole model}
\label{subsec:bipole}

Solar active regions are often approximated as bipolar magnetic regions. Their properties, including emergence latitude, spatial extension, tilt angle, and magnetic flux, have been measured for decades \citep[e.g.][]{wang1989,harveyk1993,stenflo2012}. Bipole models are commonly used as the source term for surface flux transport models \citep[e.g.][]{vanballegooijen1998,baumann2004,cameron2010}, or as the boundary condition for the background magnetic field in CME models \citep[e.g.][]{amari1996,torok2003,aulanier2010}. They have also been scaled to estimate the magnetic energy available to stellar flares \citep{aulanier2013}.

We adopt a symmetric solar bipole model \citep{yeates2020} that can be scaled to represent larger starspots. The centroids of the positive and negative polarities are placed at $(s_+,\phi_+)$ and $(s_-,\phi_-)$, respectively. Here $s$ is the sine of latitude $\lambda$, i.e. $s=\sin \lambda = \cos \theta$. The overall bipole centroid location $(s_0,\phi_0)$, the polarity separation (hereafter `bipole size') $\rho$, and the tilt angle $\gamma$ with respect to the equator are
\begin{equation}
\begin{split}
s_0 &= (s_+ + s_-) / 2, \\
\phi_0 &= (\phi_+ + \phi_-) / 2, \\
\rho &= \arccos \left[ s_+ s_- + \sqrt{1-s^2_+} \sqrt{1-s^2_-} \cos(\phi_+-\phi_-) \right],\\
\gamma &= \arctan \left[ \dfrac{\arcsin (s_+) - \arcsin (s_-)}{\sqrt{1-s^2_0} \, (\phi_- -\phi_+)} \right].\\
\end{split}
\label{eq:param}
\end{equation}
The bipole properties are defined by these parameters along with the unsigned magnetic flux $\lvert \Phi \rvert$.

To obtain the surface magnetic field map, we first rotate the spherical coordinate $(s,\phi)$ to $(s',\phi')$ such that the bipole is centred on the equator $(s'_0,\phi'_0)=(0,0)$ with zero tilt $\gamma'=0$ (Appendix A of \citealp{yeates2020}). This is equivalent to $s'_+=s'_-=0$, and $\phi'_-=-\phi'_+=\rho/2$. With the mapping from $(\theta,\phi)$ to $(s',\phi')$, the desired $B_r(R_\star,\theta,\phi)$ can be evaluated in the new coordinate system as:
\begin{equation}
F(s',\phi')= -B_0 \dfrac{\phi'}{\rho} \exp \left[ - \dfrac{\phi'^2 + 2 \arcsin^2 (s') }{(a\rho)^2} \right].
\label{eq:bipole}
\end{equation}
The constant $a$ controls the axial dipole moment of the bipole, and the positive $B_0$ controls $\lvert \Phi \rvert$.

For active regions in solar cycle 24, $a=0.56$ provides a good match with the observed dipole moment. The maximum field strength from equation~(\ref{eq:bipole}) is $aB_0/\sqrt{2e}\approx0.24B_0$.


\begin{figure}
\centerline{\includegraphics{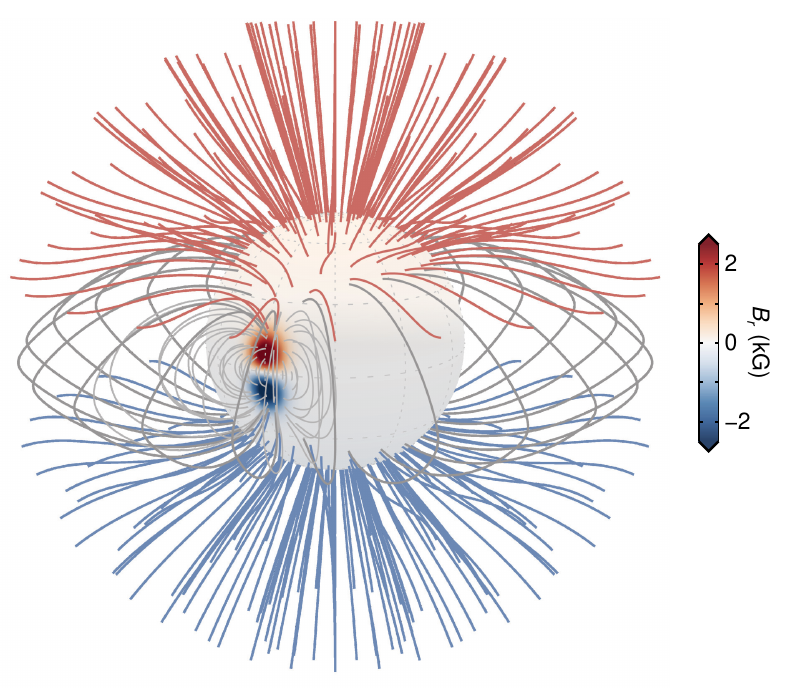}}
\caption{Illustration of the modelled coronal magnetic topology with $\rho=20\degr$, $g_{10}=200$\,G, and $R_s=2.5R_\star$. The central sphere shows the surface $B_r$. Open field lines with red (blue) colour have positive (negative) magnetic polarity. Closed field lines in dark grey have the same apex height at $r=2.4R_\star$ along the equator. Closed field lines in light grey are rooted within or near the starspots. An interactive version of the figure is available in the supplementary material and via \href{https://doi.org/10.5281/zenodo.5498419}{Zenodo}.}
\label{f:pfss}
\end{figure}


\subsection{A model for the stellar coronal magnetic field}
\label{subsec:setup}

We now consider a cool star for which the photospheric magnetic field comprises a pair of spots embedded in a global, axial dipole (Fig.~\ref{f:pfss}). The orientations of the bipole and the dipole are aligned such that the coronal field strength is maximized. The surface $B_r$ of the bipole is prescribed by equation~(\ref{eq:bipole}), whilst the dipole contribution is $g_{10}\cos\theta$. We further assume the following:
\begin{itemize}[noitemsep,topsep=2pt,parsep=2pt,leftmargin=10pt,labelwidth=10pt]

\item The bipole straddles the equator, and its dipole moment is aligned with the north-south direction. The corresponding coronal field thus points largely southward. The bipole is specified by $s_+ = -s_- = \sin(\rho/2)$, $\phi_+ = \phi_- = 0$, and $\gamma=\pi/2$. 

\item The bipole magnetic field resembles that of sunspots. We specifically adopt $a=0.56$ and $B_0=1.25\times10^4$\,G. The latter translates to a maximum field strength of $3$\,kG, typical for sunspot umbrae \citep{solanki2006}.

\item The axial dipole moment is directed northward, i.e. $g_{10}>0$. All other $g_{\ell m}$ and $h_{\ell m}$ coefficients are $0$. The corresponding coronal field points southward.
\end{itemize}

The lower boundary condition $B_r(R_\star,\theta,\phi)$ for the coronal PFSS extrapolation is calculated on a grid with $1\degr$ spatial resolution, where $B_r(R_\star,\theta,\phi)$ is the sum of the starspot bipole and global dipole fields. We truncate the series at a maximum $\ell=120$ \citep{toth2011}. The modelled coronal field has a grid size of $0.001R_\star$ in $r$.

The modelled coronal field is largely axisymmetric away from the starspots (Fig.~\ref{f:pfss}). The closed field lines lie within a same meridional plane, and the open field lines originate from the polar regions and expand super-radially. Near the starspots, however, the closed field lines zonally expand away from the centre; the open-closed field boundaries extend toward the equator.

We acknowledge that our model configuration is highly idealised. In reality, active regions are believed to emerge preferentially within the `active latitudes', which lie below $30\degr$ for the Sun but likely higher for rapid rotators according to dynamo models \citep{schuessler1992,isik2018}. Their tilt angle with respect to the equator follows Joy's law on the Sun, with a median of less than $10\degr$ \citep{hathaway2015}. Our estimates of the TSZ heights presented in Section~\ref{sec:result} therefore likely represent upper limits.

In this set-up, we assume that a possibly eruptive flux rope (not present in the model) resides above the starspots (see Fig.~\ref{f:torus}). Its axis is expected to be nearly parallel to the magnetic polarity inversion line along the equator. The apex of the flux rope axis is located at $h=r-R_\star$, $\theta=\pi/2$, and $\phi=0$. The external poloidal component of the coronal field is north-south oriented above the polarity inversion line. Hence, we adopt $B_p \equiv B_\theta(r,\pi/2,0)$ as the strapping field. The decay index $n$ in equation~(\ref{eq:n}) is then
\begin{equation}
n(r) = \left. - \dfrac{r-R_\star}{B_\theta} \dfrac{\partial B_\theta}{\partial r} \right\vert_{\theta=\pi/2,\, \phi=0}.
\label{eq:n_pfss}
\end{equation}

In what follows, we assume that the critical index $n_c$ occurs at a value of 1.5. Opting for a larger (smaller) $n_c$ will generally increase (decrease) the values of $h_c$; it is not expected to change the conclusions qualitatively.

Hereafter, the length variables will be normalised to the stellar radius $R_\star$; normalised variables will be designated by a tilde. For example, the normalised source surface radius and critical height are $\tilde{R}_s=R_s/R_\star$ and $\tilde{h}_c=h_c/R_\star$, respectively.


\begin{figure*}
\centerline{\includegraphics{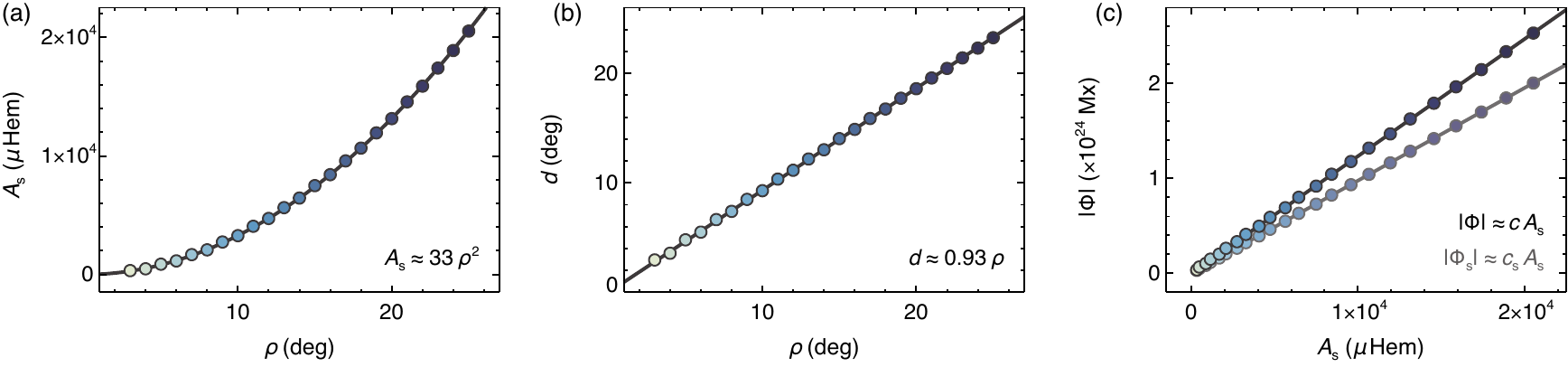}}
\caption{Surface properties of the modelled starspots, with the circles showing the results for various bipole sizes $\rho$. The curves show either a quadratic or a linear fit. The panels illustrate (a) the single spot area $A_\mathrm{s}$ as a function of $\rho$, (b) the effective spot diameter $d$ as a function of $\rho$, (c) the unsigned magnetic flux of the bipole $\lvert \Phi \rvert$ (upper branch), and within the spot pair $\lvert \Phi_{\mathrm{s}} \rvert$ (lower branch) as functions of the area of the spot pair $A_{\mathrm{s}}$. The solar radius $R_\odot=6.96\times10^{10}$\,cm is used in place of $R_\star$. The slopes are $c=1.23\times10^{20}$ and $c_\mathrm{s}=0.98\times10^{20}\,\mathrm{Mx}\,\mu\mathrm{Hem}^{-1}$. The colours in the circles are based on $\rho$ in all three panels.}
\label{f:scaling}
\end{figure*}


\begin{figure}
\centerline{\includegraphics{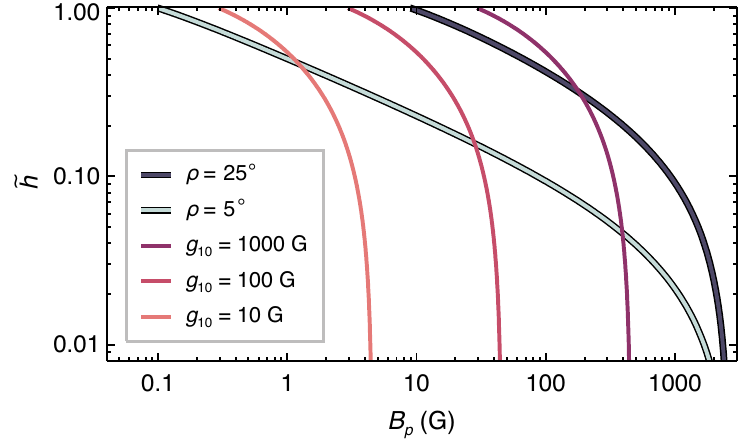}}
\caption{Height profiles of the strapping field $B_p$, for bipoles of various $\rho$ and dipoles of various $g_{10}$. The nominal $\tilde{R}_s=2.5$ is used.}
\label{f:bt}
\end{figure}


\begin{figure}
\centerline{\includegraphics{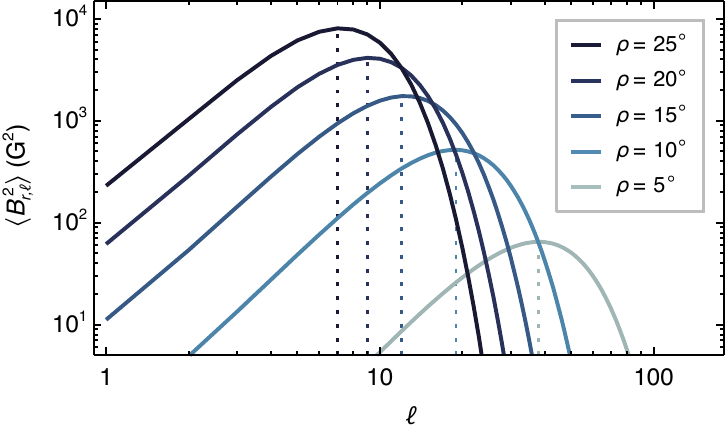}}
\caption{Energy spectra $\langle B^2_{r,\ell}\rangle$ of the surface $B_r$, as derived from equation~(\ref{eq:bl2}), for bipoles of various $\rho$. Vertical dotted lines indicate the peaks.}
\label{f:spectrum}
\end{figure}


\subsection{Free parameters}
\label{subsec:param}

Our coronal field model includes three free parameters: the bipole size ($\rho$), the axial dipole strength ($g_{10}$), and the source surface radius ($R_s$). Their values can vary significantly amongst cool stars.

\textit{Bipole size.---} Known sunspots cover at most $0.1\%$ of the solar surface area, and their sizes follow a log-normal distribution. The typical (extreme) area of a single spot is about $60$ ($3000$) microhemispheres ($\mu$Hem), or $1.3\degr$ ($8.9\degr$) in diameter \citep{baumann2005}. The average polarity separation $\rho$ of solar active regions is about $5\degr$ \citep{wang1989}. In contrast, the spot coverage of cool stars typically ranges from $1\%$ to $10\%$ based on Doppler imaging, and tends to be larger based on molecular band modelling \citep{berdyugina2005,strassmeier2009}. Recent studies using light curve modulation or occultation mapping (with transiting exoplanet) suggests that the diameter of a single spot (cluster) can reach $20\degr$ \citep{davenport2015,morris2017}.

\textit{Axial dipole strength.---} For the Sun, the axial dipole amplitude  $g_{10}$ has varied from near zero (maximum phase) to about $5$\,G (minimum phase) for the past several activity cycles \citep{derosa2012}, and remained below $10$\,G for the past century \citep{munozjaramillo2012}. Surveys show that slow-rotating, solar-mass stars and low-mass M dwarfs both tend to have poloidal, axisymmetric field, though the field amplitudes can be very different \citep{donati2009}. The estimated dipole coefficient $g_{10}$ ranges from $\log g_{10} \approx 0.3$\,G (solar-like stars) to $\log g_{10} \approx 3.2$\,G \citep[M dwarfs;][and references therein]{see2019}. The mean field $\langle B \rangle$ of inactive (active) M dwarfs are in the hectogauss (kilogauss) range \citep{kochukhov2021}. The corresponding $g_{10} \approx 2\langle B \rangle$ (assuming a pure dipolar field) is $1$--$3$ orders of magnitude stronger than the Sun.

\textit{Source surface radius.---} For the Sun, the value of $\tilde{R}_s$ may be empirically determined by comparing the PFSS modelling results with various imaging and in situ observations; $\tilde{R}_s=2.5$ is widely used \citep{hoeksema1984}. The optimal value likely varies between $1.5$ and $4$ depending on the magnetic activity level, the cycle phase, and the observable used for calibration \citep{lee2011,arden2014}. For cool stars, several scalings have been proposed. First, $\tilde{R}_s$ is required to increase with the surface magnetic activity (in the rotation-unsaturated regime) in order to reproduce the observed spin-down rate. For a star ten times more active than the Sun, the optimal $\tilde{R}_s$ is found to be about $19$ \citep{schrijver2003}. Along the same lines, a power-law scaling $\tilde{R}_s \propto P_\star^{-0.84}$ was inferred by \citet{see2018}, where $P_\star$ is the rotation period. Second, an `effective' $\tilde{R}_s$ can be derived by requiring the PFSS open magnetic flux to match ab initio stellar wind MHD models \citep{vidotto2014,reville2015}. The range $\tilde{R}_s\in[2.7,10.7]$ was found for six solar-mass stars, which increases with the surface field \citep{reville2016}. Third, $\tilde{R}_s$ should be greater than the co-rotation radius $\tilde{R}_\mathrm{co}$ so the closed field lines can support co-rotating `slingshot prominences'. For the well-studied star AB Dor with $\tilde{R}_\mathrm{co}=2.7$, $\tilde{R}_s$ was set to $3.4$ \citep{jardine2002}.

For this study, we sample this parameter space by computing a grid of models with the following parameters: $\rho \in \left[ 3\degr, 25\degr \right]$ sampled every $1\degr$; $g_{10} \in \left[ 0, 1000 \right]$\,G sampled every $10$\,G; $\tilde{R}_s=2$, $2.5$, $3$, $4$, $5$, $10$, and $20$.


\begin{figure*}
\centerline{\includegraphics{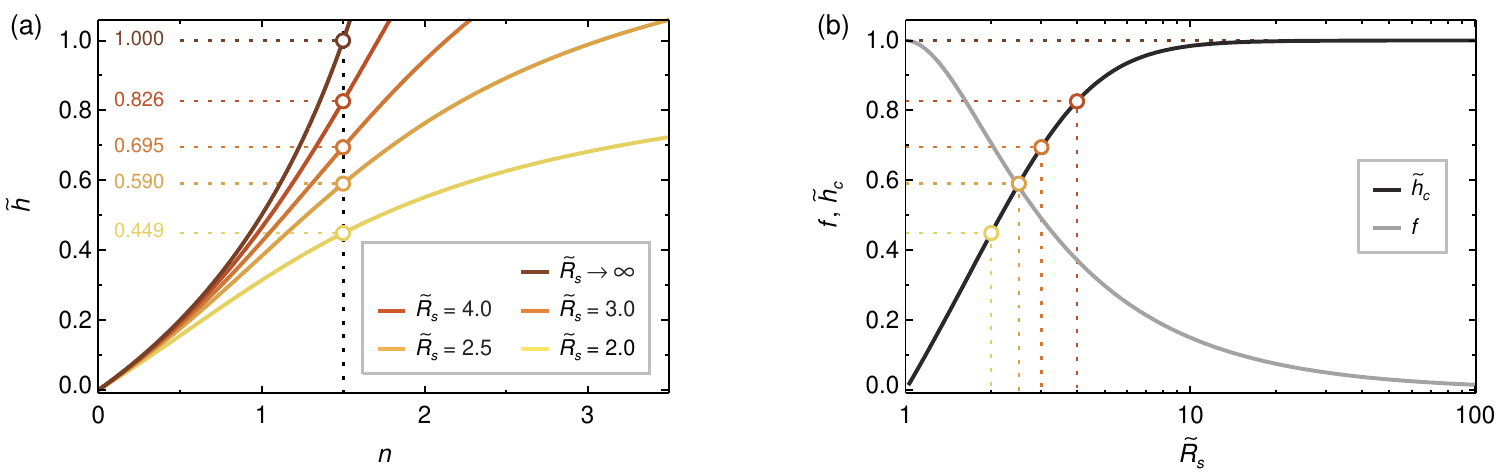}}
\caption{(a) Decay index $n$ as a function of height $\tilde{h}$ for an axial dipole field, for various source surface radii $\tilde{R}_s$. The critical decay index $n_c=1.5$ is shown as a vertical dotted line. The critical heights $\tilde{h}_c$ are indicated by circles and horizontal dotted lines. (b) Fractional open flux $f$ and $\tilde{h}_c$ as functions of $\tilde{R}_s$ for a dipole field. The cases in (a) are marked by coloured circles and dotted lines.}
\label{f:dipole}
\end{figure*}


\begin{figure*}
\centerline{\includegraphics{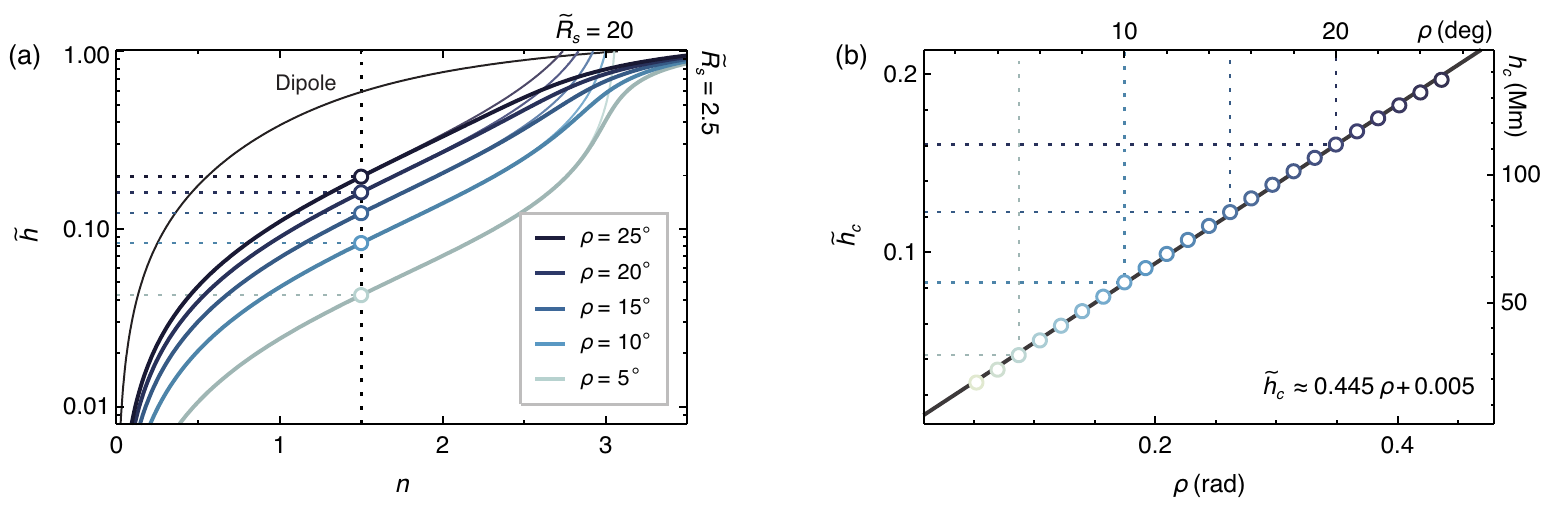}}
\caption{(a) Decay index $n$ as a function of height $\tilde{h}$ for starspots' field, for various bipole sizes $\rho$. The curves diverge into two branches at $\tilde{h}\approx0.4$, in which the lower (thicker lines) and upper (thinner lines) branches represent the cases for $\tilde{R}_s=2.5$ and $20$, respectively. The two cases are almost identical for smaller $\tilde{h}$. For comparison, the thin black curve shows the pure dipole case with $\tilde{R}_s=2.5$. The critical decay index $n_c=1.5$ is shown as a vertical dotted line. The critical heights $\tilde{h}_c$ are indicated by circles and horizontal dotted lines. (b) Critical height $\tilde{h}_c$ as a function of $\rho$ for $\tilde{R}_s=2.5$. The cases in (a) are marked by circles and dotted lines. The best fit linear function is shown. The equivalent $\rho$ in degree and $h_c$ in Mm (assuming solar radius $R_\odot=696$\,Mm) are noted on the top and right side, respectively.}
\label{f:spots}
\end{figure*}


\section{Results}
\label{sec:result}


\subsection{Model properties}
\label{subsec:property}

The properties of the modelled starspots are controlled by the bipole size $\rho$. Using a minimum penumbral $B_r$ of $700$\,G \citep{solanki2006} to define the boundary of the spots, we derive the following relations (Fig.~\ref{f:scaling}):

\begin{itemize}[noitemsep,topsep=2pt,parsep=2pt,leftmargin=10pt,labelwidth=10pt]

\item The area of a single spot is $A_{\mathrm{s}} \approx 33 \rho^2  \, \mu\mathrm{Hem}\,\mathrm{deg}^{-2}$, where $\rho$ is measured in asterographic degrees.

\item The effective angular diameter $d$ of a single spot is $d\approx0.93\rho$, assuming a circular spot whose area is $2\pi R_\star (1 - \cos(d/2))$.

\item The unsigned flux $\lvert \Phi \rvert$ of the bipole scales with the area of a single spot as $\lvert \Phi \rvert \approx c A_{\mathrm{s}}$, where $c=1.23 \left(R_\star/R_\odot\right)^2 \times10^{20}$\,Mx\,$\mu$Hem$^{-1}$.

\item The unsigned flux within the starspots is $\lvert \Phi_{\mathrm{s}} \rvert \approx c_{\mathrm{s}} A_{\mathrm{s}}$, where $c_{\mathrm{s}}=0.98 \left(R_\star/R_\odot\right)^2 \times10^{20}$\,Mx\,$\mu$Hem$^{-1}$. This accounts for $79\%$ of the bipole flux.

\end{itemize}

These empirical scalings allow for direct conversion between the bipole size and the properties of the starspots. For example, for a bipole with $\lvert \Phi \rvert = 10^{24}$\,Mx (Section~\ref{subsec:intro_confine}), $A_{\mathrm{s}}\approx8100 \times \left(R_\odot/R_\star\right)^2$\,$\mu$Hem and $d\approx14 \fdg 6 \times \left(R_\odot/R_\star\right)$. For the largest bipole considered, $\rho=25\degr$, $\lvert \Phi \rvert \approx 2.5\times  \left(R_\star/R_\odot\right)^2 \times10^{24}$\,Mx.

The decay index $n(\tilde{h})$ can be inferred from the negative slope of the $B_p(\tilde{h})$ curve in a log-log plot (Fig.~\ref{f:bt}). In our model, $B_p$ from the smaller, solar-like bipole ($\rho=5\degr$) decreases with height much faster than the largest bipole considered ($\rho=25\degr$). Therefore, the decay index of the small bipole is larger at all heights. Similarly, the decay index of the bipole's magnetic field is larger than that of the global dipole field (see Sec.~\ref{subsec:spots}).

In essence, $n(\tilde{h})$ depends on the characteristic spatial scale of the coronal field, which can be characterized by the spherical harmonic degree $\ell_\mathrm{max}$ at the peak of the energy spectrum $\langle B^2_{r,\ell} \rangle$ (Fig.~\ref{f:spectrum}). Using equation~(\ref{eq:bl2}), we find $\ell_\mathrm{max}=38$ for the small bipole with $\rho=5\degr$, in sharp contrast with $\ell_\mathrm{max}=7$ for a larger bipole with $\rho=25\degr$. Because the magnetic field components associated with larger $\ell$ decrease faster with height, their confining effect is expected to be more localised.

The critical height $\tilde{h}_c(\rho,g_{10},\tilde{R}_s)$ depends on the interplay between the spot's and the dipole's magnetic fields, which is further modulated by the source surface radius. Below, we discuss the behaviours of $n(\tilde{h})$ and $\tilde{h}_c$ for an axial dipole (Section~\ref{subsec:dipole}), for a pair of bipolar spots (Section~\ref{subsec:spots}), and for combinations of the two (Section~\ref{subsec:spdp}). The effect of the source surface is covered in each subsection.


\subsection{TSZ for an axial dipole field}
\label{subsec:dipole}

For an axial dipole field, $n$ can be calculated from equations~(\ref{eq:psi_def}) and~(\ref{eq:psi}):
\begin{equation}
n(r) = 3 \left( 1 - \dfrac{R_\star}{r} \right) \left[ 1 - \left( \dfrac{r}{R_s} \right)^3 \right]^{-1}.
\label{eq:n_g10}
\end{equation}
Plots of $n(\tilde{h})$ for various values of $\tilde{R}_s$ are shown in Fig.~5(a). The resultant $\tilde{h}_c$ is independent of $g_{10}$. Fig.~\ref{f:dipole}(b) shows that $\tilde{h}_c$ increases approximately linearly with $\log \tilde{R}_s$ for $\tilde{R}_s \lesssim 4$, and becomes almost constant for $\tilde{R}_s \gtrsim 10$. For the nominal solar value, $\tilde{R}_s=2.5$, $\tilde{h}_c=0.590$. As $\tilde{R}_s\to\infty$, which represents a closed-dipole case with no open field, we find $\tilde{h}_c = n_c / (3 - n_c) = 1$. This value of $\tilde{h}_c$ sets the upper limit of the TSZ.

By integrating $\lvert B_r \rvert$ on both the stellar photospheric surface and the source surface, one can determine both the unsigned magnetic flux $\lvert \Phi_\star \rvert$ and the open magnetic flux $\lvert \Phi_s \rvert$, respectively \citep{see2018}. The fractional open flux $f$ is
\begin{equation}
f(\tilde{R}_s)=\dfrac{\left| \Phi_s \right| }{\left| \Phi_\star \right|} = \dfrac{3 \tilde{R}_s^2 }{1 + 2 \tilde{R}_s^3},
\label{eq:phis0}
\end{equation}
which decreases with $\tilde{R}_s$ and is anti-correlated with $\tilde{h}_c$ (Fig.~\ref{f:dipole}(b)). In other words, the TSZ expands as the a larger fraction of the magnetic flux becomes more closed. For $\tilde{R}_s=2.5$, $f=0.581$.


\begin{figure*}
\centerline{\includegraphics{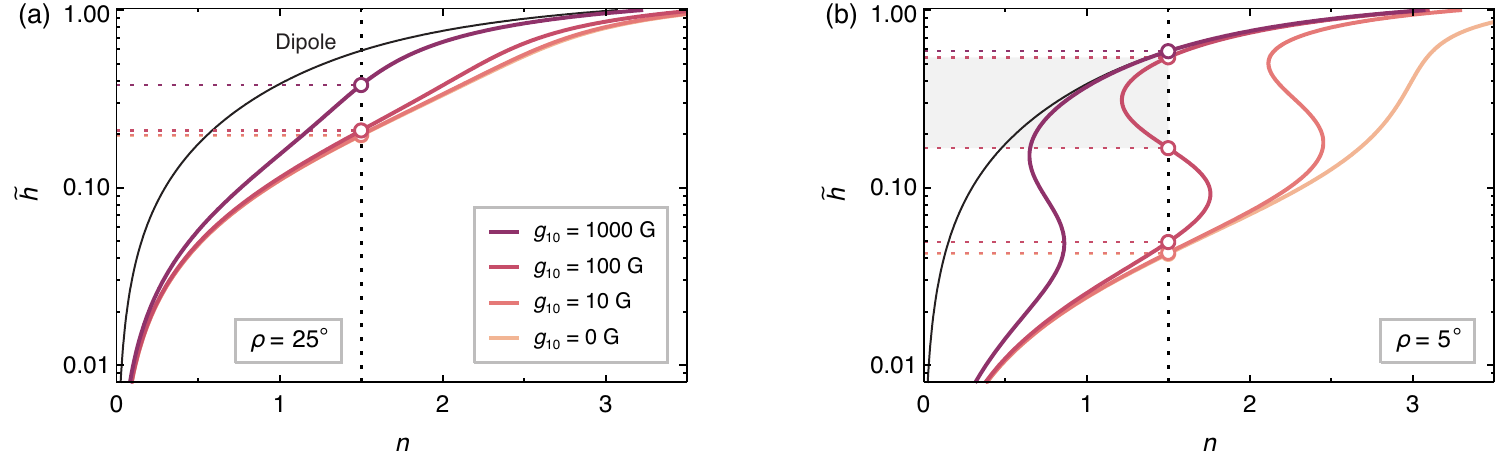}}
\caption{Decay index $n$ as a function of height $\tilde{h}$ for models with both starspots and dipole, for $\tilde{R}_s=2.5$. Panels (a) and (b) are for fixed bipole sizes $\rho=25\degr$ and $5\degr$, respectively. Coloured curves indicate various dipole strengths $g_{10}$. The thin black curve shows the dipole-only case. The critical decay index $n_c=1.5$ is shown as a vertical dotted line. The critical heights $\tilde{h}_c$ are indicated by circles and horizontal dotted lines. In (b), the $g_{10}=100$\,G curve intersects $n=n_c$ three times, with the shaded region highlighting the height range of the secondary TSZ.}
\label{f:spdp}
\end{figure*}


\begin{figure*}
\centerline{\includegraphics{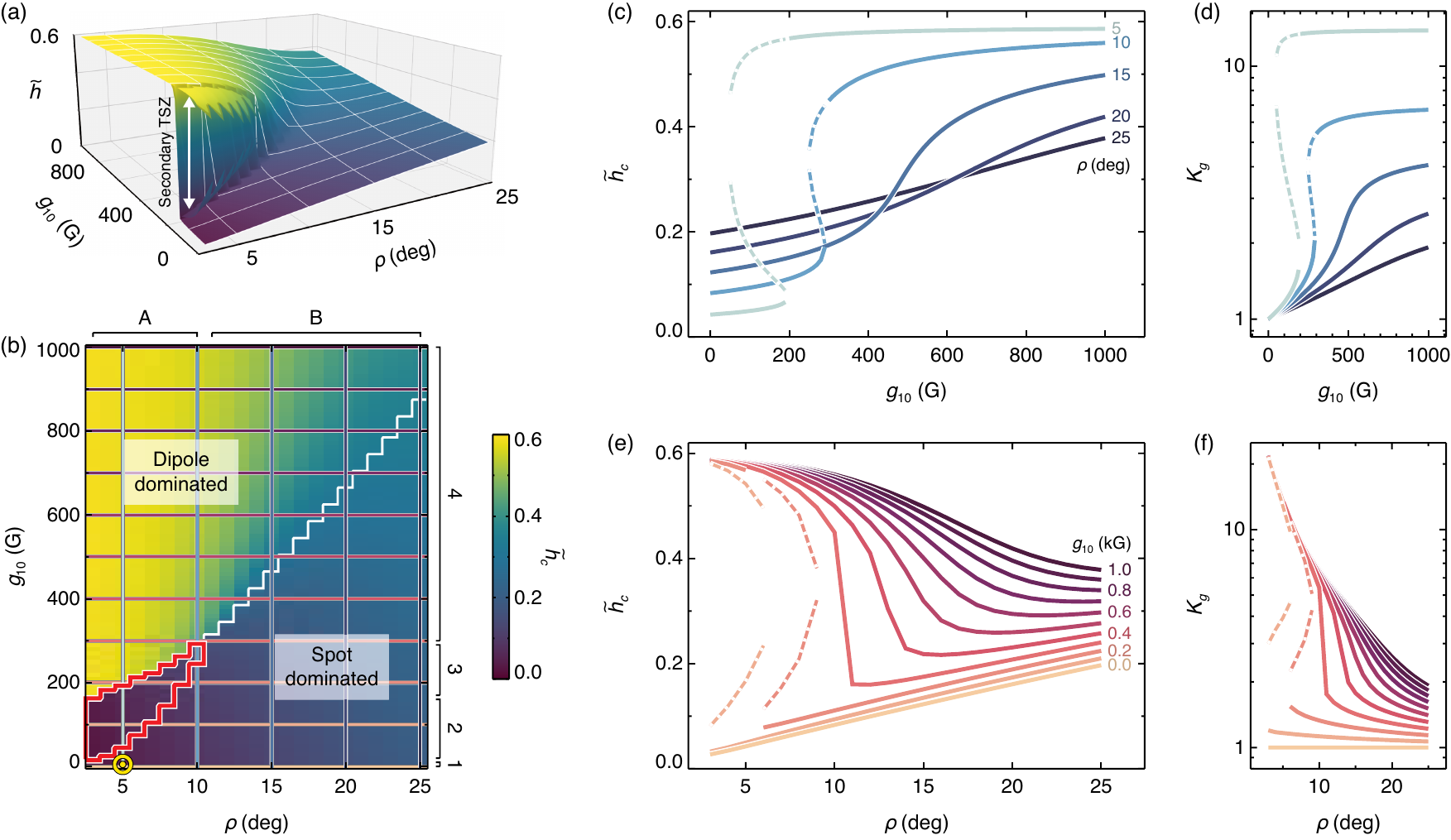}}
\caption{Extent of the TSZ for models with both starspots and a dipole, for $\tilde{R}_s=2.5$. (a) Critical height $\tilde{h}_c$ shown as a curved surface in the ($\rho$, $g_{10}$, $\tilde{h}$) space. Two surface fragments marked by white arrows outline additional values of $\tilde{h}_c$ that bracket secondary TSZs. The apparent jagged edges are due to the coarse sampling of the $\rho$-$g_{10}$ space. (b) Critical height $\tilde{h}_c(\rho,g_{10})$, shown as the background colour. The diagonal white contour indicates where the spot and dipole strapping fields $B_p$ are equal at $\tilde{h}_c$; it roughly separates the dipole- and the spot-dominated regimes. The red contour on the lower left outlines the regime where a secondary TSZ exists; the colour therein shows the lowest $\tilde{h}_c$. The typical solar value is marked by a yellow `$\odot$' symbol. The various regimes discussed in Section~\ref{subsubsec:spdp_param} are marked on the top (A and B) and on the right-hand side (1--4). (c) $\tilde{h}_c(g_{10})$, for $\rho \in \left[ 5\degr, 25\degr \right]$ sampled every $5\degr$ (vertical lines in (b)). The lower and upper dashed curves show the two values of $\tilde{h}_c$ that bracket the secondary TSZs, for $\rho=5\degr$ and $10\degr$. (d) Parameter $K_g$ (equation~(\ref{eq:k})) for the data points shown in (c). (e) $\tilde{h}_c(\rho)$, for $g_{10} \in \left[0, 1000\right]$\,G sampled every $100$\,G (horizontal lines in (b)). The dashed curves show the secondary TSZs for $g_{10}=100$ and $200$\,G. (f) Parameter $K_g$ for the data points shown in (e). An interactive version of panel (a) is available in the supplementary material and via \href{https://doi.org/10.5281/zenodo.5498419}{Zenodo}.}
\label{f:param}
\end{figure*}


\subsection{TSZ for a bipole field}
\label{subsec:spots}

The $n(\tilde{h})$ profile above a bipole (as a pair of starspots) increases more rapidly with $\tilde{h}$ compared to the dipole. The TSZ is much smaller ($\tilde{h}_c \lesssim 0.2$) and is confined to the low corona (Fig.~\ref{f:spots}(a)). Empirically, the extent of the TSZ linearly scales with the bipole size, $\tilde{h}_c \approx 0.445\rho+0.005$ (Fig.~\ref{f:spots}(b)).

The value of $\tilde{h}_c$ is largely independent of $\tilde{R}_s$. For $\tilde{R}_s=2.5$ and $20$, the relative difference is only $0.8\%$ for the largest spots considered, $\rho = 25\degr$. The effect of $\tilde{R}_s$ on $n$ manifests only higher up in the corona ($\tilde{h}\gtrsim0.4$; Fig.~\ref{f:spots}(a)).


\subsection{TSZ for starspots \& dipole}
\label{subsec:spdp}

For configurations including both starspots and a global dipole, the extension of the TSZ depends on the relative strengths of the two fields. When the bipolar starspot component predominates, this leads to a $\tilde{h}_c$ lower in the corona. In contrast, when the large-scale dipole is dominant, $\tilde{h}_c$ will be located at a greater height. We also find that the effect of the source surface to be more pronounced for dipole-dominated cases. In the following discussion, we first consider cases where $\tilde{R}_s=2.5$, and then examine the dependence on $\tilde{R}_s$ at the end of the section.

We define two non-dimensional parameters to quantify the impact of the dipole and the source surface,
\begin{equation}
\begin{split}
K_g &= {\tilde{h}_c(\rho,g_{10},\tilde{R}_s)} / {\tilde{h}_c(\rho,0,\tilde{R}_s)}, \\
K_R &= {\tilde{h}_c(\rho,g_{10},\tilde{R}_s)} / {\tilde{h}_c(\rho,g_{10},2.5)}. \\
\end{split}
\label{eq:k}
\end{equation}
The first parameter $K_g$ relates $\tilde{h}_c$ to the zero-dipole ($g_{10}=0$) case, and the second parameter $K_R$ relates $\tilde{h}_c$ to the fiducial source surface radius ($\tilde{R}_s=2.5$) case. 


\subsubsection{Two contrasting examples}
\label{subsubsec:spdp_exmp}

For the largest-spot cases with $\rho=25\degr$, $n$ monotonically increases with $\tilde{h}$ regardless of the value of $g_{10}$ (Fig.~\ref{f:spdp}(a)). For the cases with $g_{10} = 10$ and $100$\,G, the dipole contribution is relatively small (Fig.~\ref{f:bt}). The $n(\tilde{h})$ profiles therefore resemble those of the pure spots (Fig.~\ref{f:spots}), yielding similarly low TSZs: $\tilde{h}_c \le 0.210$, $K_g \le 1.07$. For $g_{10}=1000$\,G, however, the influence of the dipole field increases significantly, and the extent of the TSZ almost doubles: $\tilde{h}_c=0.378$, $K_g=1.92$.

The situation is drastically different for the smaller-spot cases with $\rho=5\degr$ owing to their fast decaying field. As $\tilde{h}$ increases, the large-scale dipole contribution quickly exceeds that of the spots, even for relatively weak dipole fields (Fig.~\ref{f:bt}). The $n(\tilde{h})$ profile is distorted into an \texttt{S} shape (Fig.~\ref{f:spdp}(b)). For $g_{10}=10$\,G, the distortion is relatively small, so the TSZ is similar to that of the spots: $\tilde{h}_c=0.043$, $K_g=1.01$. For $g_{10}=1000$\,G, however, the distortion is so large that the TSZ approaches the dipole upper bound: $\tilde{h}_c=0.586$, $K_g=13.83$ (cf. Fig.~\ref{f:bt}).

An interesting case arises for the intermediate dipole $g_{10}=100$\,G. The $n(\tilde{h})$ profile crosses $n=n_c$ three times, yielding a secondary, extended TSZ at higher altitude, $\tilde{h} \in \left[ 0.167, 0.542 \right]$. This resembles the `saddle-like' decay index profiles occasionally found on the Sun (Section~\ref{subsec:intro_ti}). The primary TSZ is similar to the TSZ of the spots, $\tilde{h}_c=0.049$, $K_g=1.17$. Its extent, $\delta \tilde{h}=0.049$, is much narrower than that of the secondary TSZ, $\delta \tilde{h}=0.375$.


\begin{figure}
\centerline{\includegraphics{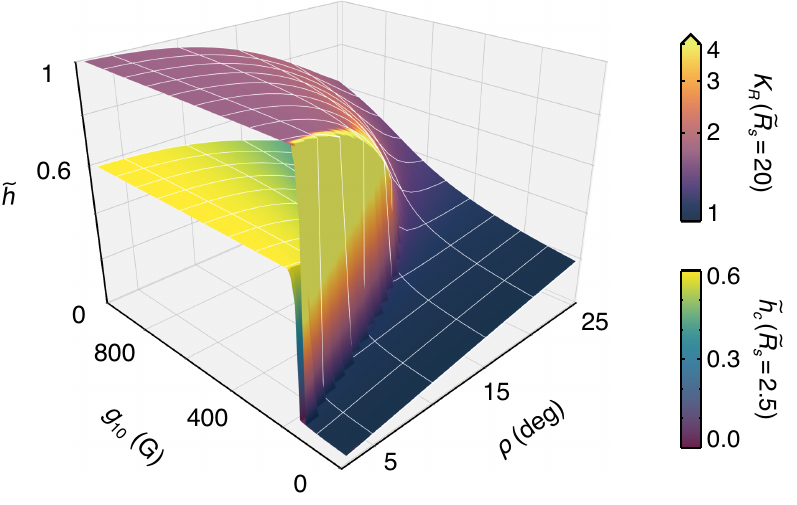}}
\caption{Response of the critical height $\tilde{h}_c$ to the source surface radius $\tilde{R}_s$. The lower surface is for $\tilde{R}_s=2.5$ (same as Fig.~\ref{f:param}(a)). The upper surface is for $\tilde{R}_s=20$, the colour of which illustrates the parameter $K_R$ (equation~(\ref{eq:k})). For regions with multiple critical heights, only the lowest is shown. An interactive version of the figure is available in the supplementary material and via \href{https://doi.org/10.5281/zenodo.5498419}{Zenodo}.}
\label{f:rss}
\end{figure}


\subsubsection{The $\rho$--$g_{10}$ space}
\label{subsubsec:spdp_param}

The function $\tilde{h}_c(\rho,g_{10})$ is quite non-linear (Fig.~\ref{f:param}(a)). Large gradients appear at the transition between the dipole- and spot-dominated regimes (upper-left vs lower-right regions in the figure; see also panel (b)), especially for the smaller spots where the influence of the dipole is stronger (larger $K_g$).

The secondary TSZ exclusively appears in the outlined region in the lower-left corner of Fig.~\ref{f:param}(b), for smaller spots $\rho \in \left[3\degr, 10\degr \right]$ and weak to intermediate dipoles $g_{10} \in \left[20, 290\right]$\,G. For the secondary TSZ to appear, the dipole needs to be strong enough to distort the $n(\tilde{h})$ profile, but not too strong such that $\tilde{h}_c$ is located too high up in the corona (see Fig.~\ref{f:spdp}(b)).

We use vertical cuts in the $(\rho,g_{10})$ space, $\tilde{h}_c(g_{10})$, to evaluate how the TSZ above starspots of a certain size varies for different dipole strengths (Fig.~\ref{f:param}(c)). Two regimes (A and B) emerge (cf. Fig.~\ref{f:param}(b)).

\setlist[enumerate,1]{label={\Alph*.}}
\begin{enumerate}[noitemsep,topsep=2pt,parsep=2pt,leftmargin=14pt,labelwidth=10pt]

\item $\rho \in \left[3\degr, 10\degr \right]$: a single $\tilde{h}_c$ initially increases with $g_{10}$ in the low corona. As $g_{10}$ continues to increase, an inflection point will be reached where three instances of $\tilde{h}_c$ arise; the higher two define a secondary TSZ that expands in size. Finally, as $g_{10}$ increases further, another inflection point is reached, above which only a single $\tilde{h}_c$ remains, located higher up in the corona.

\item $\rho \in \left[11\degr, 25\degr \right]$: $\tilde{h}_c$ increases with $g_{10}$ with no secondary TSZ. For smaller $\rho$, $\tilde{h}_c$ increases faster and reaches a greater maximum.

\end{enumerate}

The behaviour of $\tilde{h}_c(g_{10})$ in regime A can be explained in the context of Fig.~\ref{f:spdp}(b). Therein, $\tilde{h}_c$ is visualised by the intercepts between the $n(\tilde{h})$ curve and $n=n_c$. As $g_{10}$ increases, the number of intercepts increases from one to three, then reduces back to one. During the process, the higher two intercepts move apart. The lower two approach each other and finally disappear altogether.

Similarly, we use horizontal cuts $\tilde{h}_c(\rho)$ to evaluate how the TSZ varies for different spot sizes on a star with a certain dipole strength (Fig.~\ref{f:param}(e)). Four regimes ($1$--$4$) emerge (cf. Fig.~\ref{f:param}(b)).

\setlist[enumerate,1]{label={\arabic*.}}
\begin{enumerate}[noitemsep,topsep=2pt,parsep=2pt,leftmargin=12pt,labelwidth=10pt]

\item $g_{10} \in \left[0, 10\right]$\,G: $\tilde{h}_c$ linearly increases with $\rho$ in the low corona.

\item $g_{10} \in \left[20, 160\right]$\,G: a secondary TSZ appears; its extent shrinks with $\rho$. The lowest $\tilde{h}_c$ increases linearly with $\rho$.

\item $g_{10} \in \left[170, 290\right]$\,G: the profile is similar to regime 2, except only one high-lying $\tilde{h}_c$ exists for the smallest values of $\rho$.

\item $g_{10} \in \left[300, 1000\right]$\,G: the secondary TSZ completely disappears. The value of $\tilde{h}_c$ monotonically decreases with $\rho$.

\end{enumerate}

These behaviours can be broadly understood as a transition from the spot-dominated to the dipole-dominated regime. The Sun lies in regime 1, where the spots determine the extent of the TSZ. For regime-4 stars with strong dipoles, large spots act to shrink the TSZ: the inclusion of the spot field changes the slope of $B_p(\tilde{h})$. At $g_{10}=1000$\,G, $K_g$ decreases from the maximum $21.81$ for $\rho=3\degr$ to $1.92$ for $\rho=25\degr$ (Fig.~\ref{f:param}(f)).


\subsubsection{Effect of $\tilde{R}_s$}
\label{subsubsec:spdp_rs}

The behaviour of $\tilde{h}_c(\rho,g_{10})$ is qualitatively similar for all considered values of $\tilde{R}_s$. The dipole (spot) component controls the upper left (lower right) portion of the parameter space, with a sharp transition in between (analogous to the large gradient evident in Fig.~\ref{f:rss}). The effect of $\tilde{R}_s$ is more pronounced in the dipole-dominated regime, in which larger values of $\tilde{R}_s$ yield higher values of $\tilde{h}_c$, which approach the dipole upper bound. The spot-dominated regime, in contrast, remains largely unaffected by $\tilde{R}_s$. This is reflected in the $K_R$ values for $\tilde{R}_s=20$, which are approximately $1.7$ and $1$ in the dipole- and spot-dominated regime, respectively. The former number is simply the ratio of the two dipole critical heights, i.e. $0.998$ (for $\tilde{R}_s=20$) vs $0.590$ (for $\tilde{R}_s=2.5$; see Fig.~\ref{f:dipole}).

The secondary TSZ is present for all instances of $\tilde{R}_s$, similarly confined to an approximately triangular region in the lower-left corner of the $(\rho,g_{10})$ space (see Fig.~\ref{f:param}(b)). The upper-right tip of the triangle shifts to the right as $\tilde{R}_s$ increases, reaching $\rho=15\degr$ for $\tilde{R}_s=20$. As a result, the jump to higher $\tilde{h}_c$ occurs at smaller $g_{10}$ for the same $\rho$. $K_R$ becomes large at the transition region (Fig.~\ref{f:rss}).


\section{Summary \& Discussion}
\label{sec:disc}

Using an idealised coronal magnetic field model, we estimate the extent of the TSZ above a bipolar stellar active region (that is, a pair of starspots) embedded in a global dipole field. We find that the upper bound of the TSZ, defined by the torus instability critical height, $\tilde{h}_c$, increases with the spot size $\rho$, the dipole strength $g_{10}$, and the source surface radius $\tilde{R}_s$. The upper bound of the TSZ, and the presence or not of a secondary TSZ, depend on the interplay between the spots' and the dipole's magnetic fields. Our main findings are as follows.

\begin{itemize}[noitemsep,topsep=2pt,parsep=2pt,leftmargin=10pt,labelwidth=10pt]

\item For a purely dipolar field, the upper bound of the TSZ is located at a significant fraction of the stellar radius, and it is independent of the strength of the dipole.

\item  Increasing the source-surface radius will move this upper bound farther out, as the amount of closed flux increases. The upper bound of the TSZ ranges from $0.59R_\star$ for a solar-like case with ${R}_s=2.5R_\star$ to $R_\star$ for a fully closed dipole.

\item The field associated with a pair of bipolar starspots alone yields TSZ upper bounds of about half the corresponding bipole size, typically below $0.2R_\star$, i.e. much lower than the upper bounds of a pure dipole field.

\item When both a global dipole and a pair of starspots are present, the TSZ extension is determined by the relative strength of their magnetic fields. The TSZ's upper bound can change drastically at the transition between the dipole-dominated and the spot-dominated regimes, especially for smaller spots.

\item A secondary TSZ forms at a higher altitude for small spots (a few degrees) and intermediate dipoles (deca- to low hectogauss).

\end{itemize}


Active cool stars are expected to have a strong dipole field and a large source surface radius (Section~\ref{subsec:param}), both of which can significantly extend the TSZ compared to solar conditions. Assuming that CMEs on cool stars originate from torus-unstable flux ropes in the same way as on the Sun, the presence of an extended TSZ should reduce the CME occurrence above starspots. The processes facilitating the onset of the torus instability (see Section~\ref{subsec:intro_ti}) would have to operate longer, or more efficiently, to lift a flux rope out of the TSZ. Even if an eruption were driven predominantly by (very efficient) reconnection above and below a flux rope, it appears that overcoming the confining effect of a strong dipole field would remain difficult \citep{devore2008}.

The calculations presented here suggest that eruptions from small spots will be preferentially impacted. First, the $K_g$ value is large under strong dipoles (regimes A4, Fig.~\ref{f:param}(b)). The critical height $\tilde{h} \approx K_g \times 0.445 \rho$ can be an order of magnitude larger than that of the spots alone. A pre-eruption magnetic flux rope would need to be stretched into an extremely prolate shape before the torus instability can set in. The required work against the magnetic tension force makes it energetically unfavourable. Second, the presence of the secondary TSZ may lead to failed eruptions under intermediate dipoles (regimes A2 and A3). The lower, torus-unstable zone now allows for the onset of the torus instability, while the extended, higher TSZ provides a persistently strong strapping force to potentially entrap the ejecta.

The Sun's dipole field is generally too weak to produce a secondary TSZ (regime A1). In certain multipolar active regions, however, several sets of closed field lines can co-exist with very different connectivities and apex heights. The higher-lying field lines can mimic the effect the global dipole field, yielding a similar \texttt{S}-shaped decay index profile (Section~\ref{subsec:intro_ti}).


The modelled spots have a maximum surface field $B_r=3$\,kG. This value is typical for sunspot umbrae, and is consistent with the $2$--$4$\,kG estimate for active dwarf spots using the Zeeman broadening technique \citep{berdyugina2005}. A set of ab initio MHD simulations of spots on cool stars finds the umbral field to be in the range of $3$--$4.5$\,kG, which is largely determined by the fluid pressure at the unity optical depth \citep{panja2020}.

Our model yields an empirical relation between the bipole magnetic flux and the spot area. To compare with the literature, we normalise the flux as $\lvert \Phi_{21} \rvert = \lvert \Phi \rvert / (10^{21}\,\mathrm{Mx})$, and the area of the spot pair as $A_{\mathrm{s}18} = (2 A_{\mathrm{s}} \times 10^{-6} \times 2 \pi R^2_\star) / (10^{18}\,\mathrm{cm}^2\,\mu\mathrm{Hem})$. This yields $ \lvert \Phi_{21} \rvert = 2.0 A_{\mathrm{s}18}$ for $R_\star=R_\odot$, similar to the value found for the Sun, $ \lvert \Phi_{21} \rvert = 2.3 A_{\mathrm{s}18}$ \citep{wang1989}. The empirical power-law $\lvert \Phi \rvert \propto \rho^2$ is somewhat steeper than that of the Sun, $\rho^{1.3}$, possibly because the latter accounts for extended plage regions with weaker, hectogauss fields. 

The empirical spot-generated TSZ, $\tilde{h}_c\approx0.445\rho$, is similar to the $0.5\rho$ estimate from solar coronal field extrapolation models \citep{wangd2017} and idealised MHD simulations \citep{torok2007}. However, a different series of MHD simulations show that $\tilde{h}_c$ can be comparable to $\rho$ \citep{aulanier2010,zuccarello2015}. The scaling may depend on the detailed surface flux distribution, e.g. the `compactness' of the spots, or the prescribed flux rope properties.

The modelled spots are centred on the equator and aligned with the global dipole (Fig.~\ref{f:pfss}). A different, more realistic bipole centroid or orientation (Section~\ref{subsec:setup}) would reduce the dipole contribution to the strapping field in our model. Hence, the height estimates of the TSZ shown in this study should be considered as upper limits.

If a flux rope in an active region remains stable (or is even entirely absent), reconnection between sheared magnetic fields can still lead to flares without a CME \citep{lit2019}, a scenario that has been proposed to explain the observed behaviour of NOAA active region 12192 \citep{jiang2016}. This scenario may serve as another resolution to the missing stellar CME conundrum, but it remains to be seen how frequently such flares actually occur. It appears that very strong fields would be needed to produce sufficient flaring from reconnection processes that are not driven by an actual eruption.

Our model does not consider elongated quiescent filament channels, which constitute a main source for solar CMEs, especially during phases of low solar activity. Due to the weak photospheric field strength (several gauss),  eruptions stemming from such channels tend to have little flare emission. Such `flare-less' CMEs may have been overlooked on other cool stars. 

The torus instability theory is applicable in a low plasma-$\beta$ environment where the magnetic fields determine the plasma dynamics. The plasma-$\beta$ can vary significantly in the solar atmosphere, due to the changing field strength, plasma temperature, and density \citep{gary2001,bourdin2017}. Nevertheless, the $\beta \ll 1$ criterion is expected to be satisfied in the low corona above sunspots, at least up to a couple tenths of the solar radius where most active region CMEs erupt. Active stars with stronger magnetic fields are expected to have a more extended low-$\beta$ zone, so the results presented here remain relevant. 


Our idealised model is intended to provide a first-order description of the TSZ in the $(\rho,g_{10},\tilde{R}_s)$ parameter space, applicable to cool stars with a single active region. 
The coronal magnetic field configurations of magnetically more active stars are likely to be more complex. Data-constrained numerical models \citep[e.g.][]{alvaradogomez2018,lynch2019,jinm2020} will be needed to explore their effects on eruptions. In particular:

\begin{itemize}[noitemsep,topsep=2pt,parsep=2pt,leftmargin=10pt,labelwidth=10pt]

\item The model parameters are expected to be positively correlated with one another. For instance, surface flux transport simulations tailored to represent active stars yield stronger dipole fields when more large bipoles are included \citep{schrijver2001,lehmann2018}.

\item Many K and early M dwarfs have non-axisymmetric or predominantly toroidal global fields \citep[e.g.][]{donati2009}. The effect of such fields on the TSZ extension is unclear.

\item Large sunspot groups are typically fragmented \citep[see][for a discussion in the context of stellar flares]{aulanier2013}. Episodes of emergence, decay, and coalescence frequently occur. Forward modelling suggests the same (clustering and nesting) for starspots \citep{schrijver2020,isik2020}. If the spots are no longer monolithic, the presence of multiple shorter magnetic flux ropes becomes more likely than the presence a single, long rope.

\item Large polar spots with a ring of azimuthal fields are sometimes observed on rapid rotators \citep[see][and references therein]{berdyugina2005,strassmeier2009}. Surface flux transport models suggest that they form due to the accumulation of remnant active-region flux, with one polarity at the center and the opposite polarity at the periphery \citep{schrijver2001,mackay2004}. It is unclear whether the high-latitude flares observed on rapid rotators \citep{maggio2000,ilin2021} originate from such long-lived polar spots, or from newly emerged bipolar active regions.

\end{itemize}


\section*{Acknowledgements}

We thank Bernhard Kliem and Jamie Tayar for helpful discussions. X. Sun is supported by NSF award 1854760. T. T\"{o}r\"{o}k is supported by NSF award 1854790 and NASA award 80NSSC20K1274. M. L. DeRosa is supported by NASA award 80NSSC18K0029 to Lockheed Martin. Visualization softwares include the \texttt{Plotly} \textit{Python} package.


\section*{Data availability}
The interactive version of Fig.~\ref{f:pfss}, Fig.~\ref{f:param}(a), and Fig.~\ref{f:rss}, and the data sets used to generate Figs.~\ref{f:scaling}--\ref{f:rss} are available at \url{https://doi.org/10.5281/zenodo.5498419}.




\bibliographystyle{mnras}
\bibliography{TIstar}

\begin{thebibliography}{}
\makeatletter
\relax
\def\mn@urlcharsother{\let\do\@makeother \do\$\do\&\do\#\do\^\do\_\do\%\do\~}
\def\mn@doi{\begingroup\mn@urlcharsother \@ifnextchar [ {\mn@doi@}
  {\mn@doi@[]}}
\def\mn@doi@[#1]#2{\def\@tempa{#1}\ifx\@tempa\@empty \href
  {http://dx.doi.org/#2} {doi:#2}\else \href {http://dx.doi.org/#2} {#1}\fi
  \endgroup}
\def\mn@eprint#1#2{\mn@eprint@#1:#2::\@nil}
\def\mn@eprint@arXiv#1{\href {http://arxiv.org/abs/#1} {{\tt arXiv:#1}}}
\def\mn@eprint@dblp#1{\href {http://dblp.uni-trier.de/rec/bibtex/#1.xml}
  {dblp:#1}}
\def\mn@eprint@#1:#2:#3:#4\@nil{\def\@tempa {#1}\def\@tempb {#2}\def\@tempc
  {#3}\ifx \@tempc \@empty \let \@tempc \@tempb \let \@tempb \@tempa \fi \ifx
  \@tempb \@empty \def\@tempb {arXiv}\fi \@ifundefined
  {mn@eprint@\@tempb}{\@tempb:\@tempc}{\expandafter \expandafter \csname
  mn@eprint@\@tempb\endcsname \expandafter{\@tempc}}}

\bibitem[\protect\citeauthoryear{{Airapetian} et~al.,}{{Airapetian}
  et~al.}{2020}]{airapetian2020}
{Airapetian} V.~S.,  et~al., 2020, \mn@doi [Int. J. Astrobiology]
  {10.1017/S1473550419000132}, \href
  {https://ui.adsabs.harvard.edu/abs/2020IJAsB..19..136A} {19, 136}

\bibitem[\protect\citeauthoryear{{Alt} et~al.,}{{Alt} et~al.}{2021}]{alt2021}
{Alt} A.,  et~al., 2021, \mn@doi [ApJ] {10.3847/1538-4357/abda4b}, \href
  {https://ui.adsabs.harvard.edu/abs/2021ApJ...908...41A} {908, 41}

\bibitem[\protect\citeauthoryear{{Altschuler} \& {Newkirk}}{{Altschuler} \&
  {Newkirk}}{1969}]{altschuler1969}
{Altschuler} M.~D.,  {Newkirk} G.,  1969, \mn@doi [Sol. Phys.]
  {10.1007/BF00145734}, \href {https://ui.adsabs.harvard.edu/abs/1969Sol.
  Phys.....9..131A} {9, 131}

\bibitem[\protect\citeauthoryear{{Alvarado-G{\'o}mez}, {Drake}, {Cohen},
  {Moschou}  \& {Garraffo}}{{Alvarado-G{\'o}mez}
  et~al.}{2018}]{alvaradogomez2018}
{Alvarado-G{\'o}mez} J.~D.,  {Drake} J.~J.,  {Cohen} O.,  {Moschou} S.~P.,
  {Garraffo} C.,  2018, \mn@doi [ApJ] {10.3847/1538-4357/aacb7f}, \href
  {https://ui.adsabs.harvard.edu/abs/2018ApJ...862...93A} {862, 93}

\bibitem[\protect\citeauthoryear{{Amari}, {Luciani}, {Aly}  \&
  {Tagger}}{{Amari} et~al.}{1996}]{amari1996}
{Amari} T.,  {Luciani} J.~F.,  {Aly} J.~J.,   {Tagger} M.,  1996, \mn@doi
  [ApJL] {10.1086/310158}, \href
  {https://ui.adsabs.harvard.edu/abs/1996ApJ...466L..39A} {466, L39}

\bibitem[\protect\citeauthoryear{{Andrews}}{{Andrews}}{2003}]{andrews2003}
{Andrews} M.~D.,  2003, \mn@doi [Sol. Phys.]
  {10.1023/B:SOLA.0000013039.69550.bf}, \href
  {https://ui.adsabs.harvard.edu/abs/2003Sol. Phys...218..261A} {218, 261}

\bibitem[\protect\citeauthoryear{{Antiochos}, {DeVore}  \&
  {Klimchuk}}{{Antiochos} et~al.}{1999}]{antiochos1999}
{Antiochos} S.~K.,  {DeVore} C.~R.,   {Klimchuk} J.~A.,  1999, \mn@doi [ApJ]
  {10.1086/306563}, \href
  {https://ui.adsabs.harvard.edu/abs/1999ApJ...510..485A} {510, 485}

\bibitem[\protect\citeauthoryear{{Arden}, {Norton}  \& {Sun}}{{Arden}
  et~al.}{2014}]{arden2014}
{Arden} W.~M.,  {Norton} A.~A.,   {Sun} X.,  2014, \mn@doi [J. Geophys. Res.]
  {10.1002/2013JA019464}, \href
  {https://ui.adsabs.harvard.edu/abs/2014JGRA..119.1476A} {119, 1476}

\bibitem[\protect\citeauthoryear{{Argiroffi} et~al.,}{{Argiroffi}
  et~al.}{2019}]{argiroffi2019}
{Argiroffi} C.,  et~al., 2019, \mn@doi [Nat. Astron.]
  {10.1038/s41550-019-0781-4}, \href
  {https://ui.adsabs.harvard.edu/abs/2019NatAs...3..742A} {3, 742}

\bibitem[\protect\citeauthoryear{{Aulanier}, {T{\"o}r{\"o}k}, {D{\'e}moulin}
  \& {DeLuca}}{{Aulanier} et~al.}{2010}]{aulanier2010}
{Aulanier} G.,  {T{\"o}r{\"o}k} T.,  {D{\'e}moulin} P.,   {DeLuca} E.~E.,
  2010, \mn@doi [ApJ] {10.1088/0004-637X/708/1/314}, \href
  {https://ui.adsabs.harvard.edu/abs/2010ApJ...708..314A} {708, 314}

\bibitem[\protect\citeauthoryear{{Aulanier}, {D{\'e}moulin}, {Schrijver},
  {Janvier}, {Pariat}  \& {Schmieder}}{{Aulanier} et~al.}{2013}]{aulanier2013}
{Aulanier} G.,  {D{\'e}moulin} P.,  {Schrijver} C.~J.,  {Janvier} M.,  {Pariat}
  E.,   {Schmieder} B.,  2013, \mn@doi [A\&A] {10.1051/0004-6361/201220406},
  \href {https://ui.adsabs.harvard.edu/abs/2013A&A...549A..66A} {549, A66}

\bibitem[\protect\citeauthoryear{{Baker} \& {Lanzerotti}}{{Baker} \&
  {Lanzerotti}}{2016}]{baker2016}
{Baker} D.~N.,  {Lanzerotti} L.~J.,  2016, \mn@doi [Am. J. Phys.]
  {10.1119/1.4938403}, \href
  {https://ui.adsabs.harvard.edu/abs/2016AmJPh..84..166B} {84, 166}

\bibitem[\protect\citeauthoryear{{Bateman}}{{Bateman}}{1978}]{bateman1978}
{Bateman} G.,  1978, {MHD instabilities}.
Cambridge, MA: MIT Press

\bibitem[\protect\citeauthoryear{{Baumann} \& {Solanki}}{{Baumann} \&
  {Solanki}}{2005}]{baumann2005}
{Baumann} I.,  {Solanki} S.~K.,  2005, \mn@doi [A\&A]
  {10.1051/0004-6361:20053415}, \href
  {https://ui.adsabs.harvard.edu/abs/2005A&A...443.1061B} {443, 1061}

\bibitem[\protect\citeauthoryear{{Baumann}, {Schmitt}, {Sch{\"u}ssler}  \&
  {Solanki}}{{Baumann} et~al.}{2004}]{baumann2004}
{Baumann} I.,  {Schmitt} D.,  {Sch{\"u}ssler} M.,   {Solanki} S.~K.,  2004,
  \mn@doi [A\&A] {10.1051/0004-6361:20048024}, \href
  {https://ui.adsabs.harvard.edu/abs/2004A&A...426.1075B} {426, 1075}

\bibitem[\protect\citeauthoryear{{Baumgartner}, {Thalmann}  \&
  {Veronig}}{{Baumgartner} et~al.}{2018}]{baumgartner2018}
{Baumgartner} C.,  {Thalmann} J.~K.,   {Veronig} A.~M.,  2018, \mn@doi [ApJ]
  {10.3847/1538-4357/aaa243}, \href
  {https://ui.adsabs.harvard.edu/abs/2018ApJ...853..105B} {853, 105}

\bibitem[\protect\citeauthoryear{{Benz} \& {G{\"u}del}}{{Benz} \&
  {G{\"u}del}}{2010}]{benz2010}
{Benz} A.~O.,  {G{\"u}del} M.,  2010, \mn@doi [ARA\&A]
  {10.1146/annurev-astro-082708-101757}, \href
  {https://ui.adsabs.harvard.edu/abs/2010ARA&A..48..241B} {48, 241}

\bibitem[\protect\citeauthoryear{{Berdyugina}}{{Berdyugina}}{2005}]{berdyugina2005}
{Berdyugina} S.~V.,  2005, \mn@doi [Living Rev. Sol. Phys.] {10.12942/Living
  Rev. Solar Phys.-2005-8}, \href {https://ui.adsabs.harvard.edu/abs/2005Living
  Rev. Solar Phys.....2....8B} {2, 8}

\bibitem[\protect\citeauthoryear{{Bourdin}}{{Bourdin}}{2017}]{bourdin2017}
{Bourdin} P.~A.,  2017, \mn@doi [ApJL] {10.3847/2041-8213/aa9988}, \href
  {https://ui.adsabs.harvard.edu/abs/2017ApJ...850L..29B} {850, L29}

\bibitem[\protect\citeauthoryear{{Cameron}, {Jiang}, {Schmitt}  \&
  {Sch{\"u}ssler}}{{Cameron} et~al.}{2010}]{cameron2010}
{Cameron} R.~H.,  {Jiang} J.,  {Schmitt} D.,   {Sch{\"u}ssler} M.,  2010,
  \mn@doi [ApJ] {10.1088/0004-637X/719/1/264}, \href
  {https://ui.adsabs.harvard.edu/abs/2010ApJ...719..264C} {719, 264}

\bibitem[\protect\citeauthoryear{{Chen} \& {Krall}}{{Chen} \&
  {Krall}}{2003}]{chenj2003}
{Chen} J.,  {Krall} J.,  2003, \mn@doi [J. Geophys. Res.]
  {10.1029/2003JA009849}, \href
  {https://ui.adsabs.harvard.edu/abs/2003JGRA..108.1410C} {108, 1410}

\bibitem[\protect\citeauthoryear{{Chen}, {Ma}  \& {Zhang}}{{Chen}
  et~al.}{2013}]{chenhd2013}
{Chen} H.,  {Ma} S.,   {Zhang} J.,  2013, \mn@doi [ApJ]
  {10.1088/0004-637X/778/1/70}, \href
  {https://ui.adsabs.harvard.edu/abs/2013ApJ...778...70C} {778, 70}

\bibitem[\protect\citeauthoryear{{Cheng}, {Zhang}, {Ding}, {Guo}  \&
  {Su}}{{Cheng} et~al.}{2011}]{cheng2011}
{Cheng} X.,  {Zhang} J.,  {Ding} M.~D.,  {Guo} Y.,   {Su} J.~T.,  2011, \mn@doi
  [ApJ] {10.1088/0004-637X/732/2/87}, \href
  {https://ui.adsabs.harvard.edu/abs/2011ApJ...732...87C} {732, 87}

\bibitem[\protect\citeauthoryear{{Cheng}, {Zhang}, {Kliem}, {T{\"o}r{\"o}k},
  {Xing}, {Zhou}, {Inhester}  \& {Ding}}{{Cheng} et~al.}{2020}]{cheng2020}
{Cheng} X.,  {Zhang} J.,  {Kliem} B.,  {T{\"o}r{\"o}k} T.,  {Xing} C.,  {Zhou}
  Z.~J.,  {Inhester} B.,   {Ding} M.~D.,  2020, \mn@doi [ApJ]
  {10.3847/1538-4357/ab886a}, \href
  {https://ui.adsabs.harvard.edu/abs/2020ApJ...894...85C} {894, 85}

\bibitem[\protect\citeauthoryear{{Crosley} \& {Osten}}{{Crosley} \&
  {Osten}}{2018a}]{crosley2018a}
{Crosley} M.~K.,  {Osten} R.~A.,  2018a, \mn@doi [ApJ]
  {10.3847/1538-4357/aaaec2}, \href
  {https://ui.adsabs.harvard.edu/abs/2018ApJ...856...39C} {856, 39}

\bibitem[\protect\citeauthoryear{{Crosley} \& {Osten}}{{Crosley} \&
  {Osten}}{2018b}]{crosley2018b}
{Crosley} M.~K.,  {Osten} R.~A.,  2018b, \mn@doi [ApJ]
  {10.3847/1538-4357/aacf02}, \href
  {https://ui.adsabs.harvard.edu/abs/2018ApJ...862..113C} {862, 113}

\bibitem[\protect\citeauthoryear{{Davenport}}{{Davenport}}{2016}]{davenport2016}
{Davenport} J. R.~A.,  2016, \mn@doi [ApJ] {10.3847/0004-637X/829/1/23}, \href
  {https://ui.adsabs.harvard.edu/abs/2016ApJ...829...23D} {829, 23}

\bibitem[\protect\citeauthoryear{{Davenport}, {Hebb}  \& {Hawley}}{{Davenport}
  et~al.}{2015}]{davenport2015}
{Davenport} J. R.~A.,  {Hebb} L.,   {Hawley} S.~L.,  2015, \mn@doi [ApJ]
  {10.1088/0004-637X/806/2/212}, \href
  {https://ui.adsabs.harvard.edu/abs/2015ApJ...806..212D} {806, 212}

\bibitem[\protect\citeauthoryear{{DeRosa}, {Brun}  \& {Hoeksema}}{{DeRosa}
  et~al.}{2012}]{derosa2012}
{DeRosa} M.~L.,  {Brun} A.~S.,   {Hoeksema} J.~T.,  2012, \mn@doi [ApJ]
  {10.1088/0004-637X/757/1/96}, \href
  {https://ui.adsabs.harvard.edu/abs/2012ApJ...757...96D} {757, 96}

\bibitem[\protect\citeauthoryear{{DeVore} \& {Antiochos}}{{DeVore} \&
  {Antiochos}}{2008}]{devore2008}
{DeVore} C.~R.,  {Antiochos} S.~K.,  2008, \mn@doi [ApJ] {10.1086/588011},
  \href {https://ui.adsabs.harvard.edu/abs/2008ApJ...680..740D} {680, 740}

\bibitem[\protect\citeauthoryear{{Donati} \& {Brown}}{{Donati} \&
  {Brown}}{1997}]{donati1997}
{Donati} J.~F.,  {Brown} S.~F.,  1997, A\&A, \href
  {https://ui.adsabs.harvard.edu/abs/1997A&A...326.1135D} {326, 1135}

\bibitem[\protect\citeauthoryear{{Donati} \& {Landstreet}}{{Donati} \&
  {Landstreet}}{2009}]{donati2009}
{Donati} J.~F.,  {Landstreet} J.~D.,  2009, \mn@doi [ARA\&A]
  {10.1146/annurev-astro-082708-101833}, \href
  {https://ui.adsabs.harvard.edu/abs/2009ARA&A..47..333D} {47, 333}

\bibitem[\protect\citeauthoryear{{Drake}, {Cohen}, {Yashiro}  \&
  {Gopalswamy}}{{Drake} et~al.}{2013}]{drake2013}
{Drake} J.~J.,  {Cohen} O.,  {Yashiro} S.,   {Gopalswamy} N.,  2013, \mn@doi
  [ApJ] {10.1088/0004-637X/764/2/170}, \href
  {https://ui.adsabs.harvard.edu/abs/2013ApJ...764..170D} {764, 170}

\bibitem[\protect\citeauthoryear{{Drake}, {Cohen}, {Garraffo}  \&
  {Kashyap}}{{Drake} et~al.}{2016}]{drake2016}
{Drake} J.~J.,  {Cohen} O.,  {Garraffo} C.,   {Kashyap} V.,  2016, in
  {Kosovichev} A.~G.,  {Hawley} S.~L.,   {Heinzel} P.,  eds,  IAU Symposium
  Vol. 320, Solar and Stellar Flares and their Effects on Planets. pp 196--201
  (\mn@eprint {arXiv} {1610.05185}), \mn@doi{10.1017/S1743921316000260}

\bibitem[\protect\citeauthoryear{{Duan}, {Jiang}, {He}, {Feng}, {Zou}  \&
  {Cui}}{{Duan} et~al.}{2019}]{duan2019}
{Duan} A.,  {Jiang} C.,  {He} W.,  {Feng} X.,  {Zou} P.,   {Cui} J.,  2019,
  \mn@doi [ApJ] {10.3847/1538-4357/ab3e33}, \href
  {https://ui.adsabs.harvard.edu/abs/2019ApJ...884...73D} {884, 73}

\bibitem[\protect\citeauthoryear{{Emslie} et~al.,}{{Emslie}
  et~al.}{2012}]{emslie2012}
{Emslie} A.~G.,  et~al., 2012, \mn@doi [ApJ] {10.1088/0004-637X/759/1/71},
  \href {https://ui.adsabs.harvard.edu/abs/2012ApJ...759...71E} {759, 71}

\bibitem[\protect\citeauthoryear{{Fan}}{{Fan}}{2010}]{fan2010}
{Fan} Y.,  2010, \mn@doi [ApJ] {10.1088/0004-637X/719/1/728}, \href
  {https://ui.adsabs.harvard.edu/abs/2010ApJ...719..728F} {719, 728}

\bibitem[\protect\citeauthoryear{{Farrish}, {Alexander}, {Maruo}, {DeRosa},
  {Toffoletto}  \& {Sciola}}{{Farrish} et~al.}{2019}]{farrish2019}
{Farrish} A.~O.,  {Alexander} D.,  {Maruo} M.,  {DeRosa} M.,  {Toffoletto} F.,
   {Sciola} A.~M.,  2019, \mn@doi [ApJ] {10.3847/1538-4357/ab4652}, \href
  {https://ui.adsabs.harvard.edu/abs/2019ApJ...885...51F} {885, 51}

\bibitem[\protect\citeauthoryear{{Filippov}}{{Filippov}}{2020}]{filippov2020}
{Filippov} B.,  2020, \mn@doi [MNRAS] {10.1093/mnras/staa896}, \href
  {https://ui.adsabs.harvard.edu/abs/2020MNRAS.494.2166F} {494, 2166}

\bibitem[\protect\citeauthoryear{{Filippov} \& {Den}}{{Filippov} \&
  {Den}}{2001}]{filippov2001}
{Filippov} B.~P.,  {Den} O.~G.,  2001, \mn@doi [J. Geophys. Res.]
  {10.1029/2000JA004002}, \href
  {https://ui.adsabs.harvard.edu/abs/2001JGR...10625177F} {106, 25177}

\bibitem[\protect\citeauthoryear{{Folsom} et~al.,}{{Folsom}
  et~al.}{2016}]{folsom2016}
{Folsom} C.~P.,  et~al., 2016, \mn@doi [MNRAS] {10.1093/mnras/stv2924}, \href
  {https://ui.adsabs.harvard.edu/abs/2016MNRAS.457..580F} {457, 580}

\bibitem[\protect\citeauthoryear{{Gary}}{{Gary}}{2001}]{gary2001}
{Gary} G.~A.,  2001, \mn@doi [Sol. Phys.] {10.1023/A:1012722021820}, \href
  {https://ui.adsabs.harvard.edu/abs/2001SoPh..203...71G} {203, 71}

\bibitem[\protect\citeauthoryear{{Green}, {Kliem}, {T{\"o}r{\"o}k}, {van
  Driel-Gesztelyi}  \& {Attrill}}{{Green} et~al.}{2007}]{green2007}
{Green} L.~M.,  {Kliem} B.,  {T{\"o}r{\"o}k} T.,  {van Driel-Gesztelyi} L.,
  {Attrill} G.~D.~R.,  2007, \mn@doi [Sol. Phys.] {10.1007/s11207-007-9061-z},
  \href {https://ui.adsabs.harvard.edu/abs/2007Sol. Phys...246..365G} {246,
  365}

\bibitem[\protect\citeauthoryear{{Green}, {Kliem}  \& {Wallace}}{{Green}
  et~al.}{2011}]{green2011}
{Green} L.~M.,  {Kliem} B.,   {Wallace} A.~J.,  2011, \mn@doi [A\&A]
  {10.1051/0004-6361/201015146}, \href
  {https://ui.adsabs.harvard.edu/abs/2011A&A...526A...2G} {526, A2}

\bibitem[\protect\citeauthoryear{{Green}, {T{\"o}r{\"o}k}, {Vr{\v{s}}nak},
  {Manchester}  \& {Veronig}}{{Green} et~al.}{2018}]{green2018}
{Green} L.~M.,  {T{\"o}r{\"o}k} T.,  {Vr{\v{s}}nak} B.,  {Manchester} W.,
  {Veronig} A.,  2018, \mn@doi [Space Sci. Rev.] {10.1007/s11214-017-0462-5},
  \href {https://ui.adsabs.harvard.edu/abs/2018SSRv..214...46G} {214, 46}

\bibitem[\protect\citeauthoryear{{G{\"u}nther} et~al.,}{{G{\"u}nther}
  et~al.}{2020}]{gunther2020}
{G{\"u}nther} M.~N.,  et~al., 2020, \mn@doi [AJ] {10.3847/1538-3881/ab5d3a},
  \href {https://ui.adsabs.harvard.edu/abs/2020AJ....159...60G} {159, 60}

\bibitem[\protect\citeauthoryear{{Guo}, {Ding}, {Schmieder}, {Li},
  {T{\"o}r{\"o}k}  \& {Wiegelmann}}{{Guo} et~al.}{2010}]{guo2010}
{Guo} Y.,  {Ding} M.~D.,  {Schmieder} B.,  {Li} H.,  {T{\"o}r{\"o}k} T.,
  {Wiegelmann} T.,  2010, \mn@doi [ApJL] {10.1088/2041-8205/725/1/L38}, \href
  {https://ui.adsabs.harvard.edu/abs/2010ApJ...725L..38G} {725, L38}

\bibitem[\protect\citeauthoryear{{Harvey} \& {Zwaan}}{{Harvey} \&
  {Zwaan}}{1993}]{harveyk1993}
{Harvey} K.~L.,  {Zwaan} C.,  1993, \mn@doi [Sol. Phys.] {10.1007/BF00675537},
  \href {https://ui.adsabs.harvard.edu/abs/1993Sol. Phys...148...85H} {148, 85}

\bibitem[\protect\citeauthoryear{{Hassanin} \& {Kliem}}{{Hassanin} \&
  {Kliem}}{2016}]{hassanin2016}
{Hassanin} A.,  {Kliem} B.,  2016, \mn@doi [ApJ] {10.3847/0004-637X/832/2/106},
  \href {https://ui.adsabs.harvard.edu/abs/2016ApJ...832..106H} {832, 106}

\bibitem[\protect\citeauthoryear{{Hathaway}}{{Hathaway}}{2015}]{hathaway2015}
{Hathaway} D.~H.,  2015, \mn@doi [Living Rev. Sol. Phys.]
  {10.1007/lrsp-2015-4}, \href
  {https://ui.adsabs.harvard.edu/abs/2015LRSP...12....4H} {12, 4}

\bibitem[\protect\citeauthoryear{{Hoeksema}}{{Hoeksema}}{1984}]{hoeksema1984}
{Hoeksema} J.~T.,  1984, PhD thesis, Stanford Univ., CA.

\bibitem[\protect\citeauthoryear{{Howard}, {Corbett}, {Law}, {Ratzloff},
  {Glazier}, {Fors}, {del Ser}  \& {Haislip}}{{Howard}
  et~al.}{2019}]{howard2019}
{Howard} W.~S.,  {Corbett} H.,  {Law} N.~M.,  {Ratzloff} J.~K.,  {Glazier} A.,
  {Fors} O.,  {del Ser} D.,   {Haislip} J.,  2019, \mn@doi [ApJ]
  {10.3847/1538-4357/ab2767}, \href
  {https://ui.adsabs.harvard.edu/abs/2019ApJ...881....9H} {881, 9}

\bibitem[\protect\citeauthoryear{{Huang}, {Cheng}  \& {Ding}}{{Huang}
  et~al.}{2020}]{huangzw2020}
{Huang} Z.~W.,  {Cheng} X.,   {Ding} M.~D.,  2020, \mn@doi [ApJL]
  {10.3847/2041-8213/abc5b0}, \href
  {https://ui.adsabs.harvard.edu/abs/2020ApJ...904L...2H} {904, L2}

\bibitem[\protect\citeauthoryear{{I{\c{s}}{\i}k}, {Solanki}, {Krivova}  \&
  {Shapiro}}{{I{\c{s}}{\i}k} et~al.}{2018}]{isik2018}
{I{\c{s}}{\i}k} E.,  {Solanki} S.~K.,  {Krivova} N.~A.,   {Shapiro} A.~I.,
  2018, \mn@doi [A\&A] {10.1051/0004-6361/201833393}, \href
  {https://ui.adsabs.harvard.edu/abs/2018A&A...620A.177I} {620, A177}

\bibitem[\protect\citeauthoryear{{I{\c{s}}{\i}k}, {Shapiro}, {Solanki}  \&
  {Krivova}}{{I{\c{s}}{\i}k} et~al.}{2020}]{isik2020}
{I{\c{s}}{\i}k} E.,  {Shapiro} A.~I.,  {Solanki} S.~K.,   {Krivova} N.~A.,
  2020, \mn@doi [ApJL] {10.3847/2041-8213/abb409}, \href
  {https://ui.adsabs.harvard.edu/abs/2020ApJ...901L..12I} {901, L12}

\bibitem[\protect\citeauthoryear{{Ilin} et~al.,}{{Ilin}
  et~al.}{2021}]{ilin2021}
{Ilin} E.,  et~al., 2021, \mn@doi [MNRAS] {10.1093/mnras/stab2159}, \href
  {https://ui.adsabs.harvard.edu/abs/2021MNRAS.507.1723I} {507, 1723}

\bibitem[\protect\citeauthoryear{{Isenberg} \& {Forbes}}{{Isenberg} \&
  {Forbes}}{2007}]{isenberg2007}
{Isenberg} P.~A.,  {Forbes} T.~G.,  2007, \mn@doi [ApJ] {10.1086/522025}, \href
  {https://ui.adsabs.harvard.edu/abs/2007ApJ...670.1453I} {670, 1453}

\bibitem[\protect\citeauthoryear{{Jardine}, {Collier Cameron}  \&
  {Donati}}{{Jardine} et~al.}{2002}]{jardine2002}
{Jardine} M.,  {Collier Cameron} A.,   {Donati} J.~F.,  2002, \mn@doi [MNRAS]
  {10.1046/j.1365-8711.2002.05394.x}, \href
  {https://ui.adsabs.harvard.edu/abs/2002MNRAS.333..339J} {333, 339}

\bibitem[\protect\citeauthoryear{{Ji}, {Wang}, {Schmahl}, {Moon}  \&
  {Jiang}}{{Ji} et~al.}{2003}]{ji2003}
{Ji} H.,  {Wang} H.,  {Schmahl} E.~J.,  {Moon} Y.-J.,   {Jiang} Y.,  2003,
  \mn@doi [ApJL] {10.1086/378178}, \href
  {http://adsabs.harvard.edu/abs/2003ApJ...595L.135J} {595, L135}

\bibitem[\protect\citeauthoryear{{Jiang}, {Wu}, {Yurchyshyn}, {Wang}, {Feng}
  \& {Hu}}{{Jiang} et~al.}{2016}]{jiang2016}
{Jiang} C.,  {Wu} S.~T.,  {Yurchyshyn} V.,  {Wang} H.,  {Feng} X.,   {Hu} Q.,
  2016, \mn@doi [ApJ] {10.3847/0004-637X/828/1/62}, \href
  {https://ui.adsabs.harvard.edu/abs/2016ApJ...828...62J} {828, 62}

\bibitem[\protect\citeauthoryear{{Jin} et~al.,}{{Jin} et~al.}{2020}]{jinm2020}
{Jin} M.,  et~al., 2020, in {Kosovichev} A.,  {Strassmeier} S.,   {Jardine} M.,
   eds,  IAU Symposium Vol. 354, Solar and Stellar Magnetic Fields: Origins and
  Manifestations. pp 426--432 (\mn@eprint {arXiv} {2002.06249}),
  \mn@doi{10.1017/S1743921320000575}

\bibitem[\protect\citeauthoryear{{Jing}, {Liu}, {Lee}, {Ji}, {Liu}, {Xu}  \&
  {Wang}}{{Jing} et~al.}{2018}]{jing2018}
{Jing} J.,  {Liu} C.,  {Lee} J.,  {Ji} H.,  {Liu} N.,  {Xu} Y.,   {Wang} H.,
  2018, \mn@doi [ApJ] {10.3847/1538-4357/aad6e4}, \href
  {https://ui.adsabs.harvard.edu/abs/2018ApJ...864..138J} {864, 138}

\bibitem[\protect\citeauthoryear{{Karpen}, {Antiochos}  \& {DeVore}}{{Karpen}
  et~al.}{2012}]{karpen2012}
{Karpen} J.~T.,  {Antiochos} S.~K.,   {DeVore} C.~R.,  2012, \mn@doi [ApJ]
  {10.1088/0004-637X/760/1/81}, \href
  {https://ui.adsabs.harvard.edu/abs/2012ApJ...760...81K} {760, 81}

\bibitem[\protect\citeauthoryear{{Khodachenko} et~al.,}{{Khodachenko}
  et~al.}{2007}]{khodachenko2007}
{Khodachenko} M.~L.,  et~al., 2007, \mn@doi [Astrobiology]
  {10.1089/ast.2006.0127}, \href
  {https://ui.adsabs.harvard.edu/abs/2007AsBio...7..167K} {7, 167}

\bibitem[\protect\citeauthoryear{{Kliem} \& {T{\"o}r{\"o}k}}{{Kliem} \&
  {T{\"o}r{\"o}k}}{2006}]{kliem2006}
{Kliem} B.,  {T{\"o}r{\"o}k} T.,  2006, \mn@doi [Phys. Rev. Lett.]
  {10.1103/PhysRevLett.96.255002}, \href
  {https://ui.adsabs.harvard.edu/abs/2006PhRvL..96y5002K} {96, 255002}

\bibitem[\protect\citeauthoryear{{Kliem}, {Lin}, {Forbes}, {Priest}  \&
  {T{\"o}r{\"o}k}}{{Kliem} et~al.}{2014}]{kliem2014}
{Kliem} B.,  {Lin} J.,  {Forbes} T.~G.,  {Priest} E.~R.,   {T{\"o}r{\"o}k} T.,
  2014, \mn@doi [ApJ] {10.1088/0004-637X/789/1/46}, \href
  {https://ui.adsabs.harvard.edu/abs/2014ApJ...789...46K} {789, 46}

\bibitem[\protect\citeauthoryear{{Kliem}, {Lee}, {Liu}, {White}, {Liu}  \&
  {Masuda}}{{Kliem} et~al.}{2021}]{kliem2021}
{Kliem} B.,  {Lee} J.,  {Liu} R.,  {White} S.~M.,  {Liu} C.,   {Masuda} S.,
  2021, \mn@doi [ApJ] {10.3847/1538-4357/abda37}, \href
  {https://ui.adsabs.harvard.edu/abs/2021ApJ...909...91K} {909, 91}

\bibitem[\protect\citeauthoryear{{Kochukhov}}{{Kochukhov}}{2021}]{kochukhov2021}
{Kochukhov} O.,  2021, \mn@doi [A\&ARv] {10.1007/s00159-020-00130-3}, \href
  {https://ui.adsabs.harvard.edu/abs/2021A&ARv..29....1K} {29, 1}

\bibitem[\protect\citeauthoryear{{Lee}, {Luhmann}, {Hoeksema}, {Sun}, {Arge}
  \& {de Pater}}{{Lee} et~al.}{2011}]{lee2011}
{Lee} C.~O.,  {Luhmann} J.~G.,  {Hoeksema} J.~T.,  {Sun} X.,  {Arge} C.~N.,
  {de Pater} I.,  2011, \mn@doi [Sol. Phys.] {10.1007/s11207-010-9699-9}, \href
  {https://ui.adsabs.harvard.edu/abs/2011Sol. Phys...269..367L} {269, 367}

\bibitem[\protect\citeauthoryear{{Lehmann}, {Jardine}, {Mackay}  \&
  {Vidotto}}{{Lehmann} et~al.}{2018}]{lehmann2018}
{Lehmann} L.~T.,  {Jardine} M.~M.,  {Mackay} D.~H.,   {Vidotto} A.~A.,  2018,
  \mn@doi [MNRAS] {10.1093/mnras/sty1230}, \href
  {https://ui.adsabs.harvard.edu/abs/2018MNRAS.478.4390L} {478, 4390}

\bibitem[\protect\citeauthoryear{{Leitzinger} et~al.,}{{Leitzinger}
  et~al.}{2020}]{leitzinger2020}
{Leitzinger} M.,  et~al., 2020, \mn@doi [MNRAS] {10.1093/mnras/staa504}, \href
  {https://ui.adsabs.harvard.edu/abs/2020MNRAS.493.4570L} {493, 4570}

\bibitem[\protect\citeauthoryear{{Li}, {Liu}, {Hou}  \& {Zhang}}{{Li}
  et~al.}{2019}]{lit2019}
{Li} T.,  {Liu} L.,  {Hou} Y.,   {Zhang} J.,  2019, \mn@doi [ApJ]
  {10.3847/1538-4357/ab3121}, \href
  {https://ui.adsabs.harvard.edu/abs/2019ApJ...881..151L} {881, 151}

\bibitem[\protect\citeauthoryear{{Li}, {Hou}, {Yang}, {Zhang}, {Liu}  \&
  {Veronig}}{{Li} et~al.}{2020}]{lit2020}
{Li} T.,  {Hou} Y.,  {Yang} S.,  {Zhang} J.,  {Liu} L.,   {Veronig} A.~M.,
  2020, \mn@doi [ApJ] {10.3847/1538-4357/aba6ef}, \href
  {https://ui.adsabs.harvard.edu/abs/2020ApJ...900..128L} {900, 128}

\bibitem[\protect\citeauthoryear{{Li}, {Chen}, {Hou}, {Veronig}, {Yang}  \&
  {Zhang}}{{Li} et~al.}{2021}]{lit2021}
{Li} T.,  {Chen} A.,  {Hou} Y.,  {Veronig} A.~M.,  {Yang} S.,   {Zhang} J.,
  2021, \mn@doi [ApJL] {10.3847/2041-8213/ac1a15}, \href
  {https://ui.adsabs.harvard.edu/abs/2021ApJ...917L..29L} {917, L29}

\bibitem[\protect\citeauthoryear{{Liu}}{{Liu}}{2008}]{liuy2008}
{Liu} Y.,  2008, \mn@doi [ApJL] {10.1086/589282}, \href
  {https://ui.adsabs.harvard.edu/abs/2008ApJ...679L.151L} {679, L151}

\bibitem[\protect\citeauthoryear{{Lynch}, {Airapetian}, {DeVore}, {Kazachenko},
  {L{\"u}ftinger}, {Kochukhov}, {Ros{\'e}n}  \& {Abbett}}{{Lynch}
  et~al.}{2019}]{lynch2019}
{Lynch} B.~J.,  {Airapetian} V.~S.,  {DeVore} C.~R.,  {Kazachenko} M.~D.,
  {L{\"u}ftinger} T.,  {Kochukhov} O.,  {Ros{\'e}n} L.,   {Abbett} W.~P.,
  2019, \mn@doi [ApJ] {10.3847/1538-4357/ab287e}, \href
  {https://ui.adsabs.harvard.edu/abs/2019ApJ...880...97L} {880, 97}

\bibitem[\protect\citeauthoryear{{Mackay}, {Jardine}, {Collier Cameron},
  {Donati}  \& {Hussain}}{{Mackay} et~al.}{2004}]{mackay2004}
{Mackay} D.~H.,  {Jardine} M.,  {Collier Cameron} A.,  {Donati} J.~F.,
  {Hussain} G.~A.~J.,  2004, \mn@doi [MNRAS]
  {10.1111/j.1365-2966.2004.08233.x}, \href
  {https://ui.adsabs.harvard.edu/abs/2004MNRAS.354..737M} {354, 737}

\bibitem[\protect\citeauthoryear{{Maehara} et~al.,}{{Maehara}
  et~al.}{2012}]{maehara2012}
{Maehara} H.,  et~al., 2012, \mn@doi [Nature] {10.1038/nature11063}, \href
  {https://ui.adsabs.harvard.edu/abs/2012Natur.485..478M} {485, 478}

\bibitem[\protect\citeauthoryear{{Maggio}, {Pallavicini}, {Reale}  \&
  {Tagliaferri}}{{Maggio} et~al.}{2000}]{maggio2000}
{Maggio} A.,  {Pallavicini} R.,  {Reale} F.,   {Tagliaferri} G.,  2000, A\&A,
  \href {https://ui.adsabs.harvard.edu/abs/2000A&A...356..627M} {356, 627}

\bibitem[\protect\citeauthoryear{{Marsden} et~al.,}{{Marsden}
  et~al.}{2014}]{marsden2014}
{Marsden} S.~C.,  et~al., 2014, \mn@doi [MNRAS] {10.1093/mnras/stu1663}, \href
  {https://ui.adsabs.harvard.edu/abs/2014MNRAS.444.3517M} {444, 3517}

\bibitem[\protect\citeauthoryear{{McCauley}, {Su}, {Schanche}, {Evans}, {Su},
  {McKillop}  \& {Reeves}}{{McCauley} et~al.}{2015}]{mccauley2015}
{McCauley} P.~I.,  {Su} Y.~N.,  {Schanche} N.,  {Evans} K.~E.,  {Su} C.,
  {McKillop} S.,   {Reeves} K.~K.,  2015, \mn@doi [Sol. Phys.]
  {10.1007/s11207-015-0699-7}, \href
  {https://ui.adsabs.harvard.edu/abs/2015Sol. Phys...290.1703M} {290, 1703}

\bibitem[\protect\citeauthoryear{{Moore}, {Sterling}, {Hudson}  \&
  {Lemen}}{{Moore} et~al.}{2001}]{moore2001}
{Moore} R.~L.,  {Sterling} A.~C.,  {Hudson} H.~S.,   {Lemen} J.~R.,  2001,
  \mn@doi [ApJ] {10.1086/320559}, \href
  {https://ui.adsabs.harvard.edu/abs/2001ApJ...552..833M} {552, 833}

\bibitem[\protect\citeauthoryear{{Morris}, {Hebb}, {Davenport}, {Rohn}  \&
  {Hawley}}{{Morris} et~al.}{2017}]{morris2017}
{Morris} B.~M.,  {Hebb} L.,  {Davenport} J. R.~A.,  {Rohn} G.,   {Hawley}
  S.~L.,  2017, \mn@doi [ApJ] {10.3847/1538-4357/aa8555}, \href
  {https://ui.adsabs.harvard.edu/abs/2017ApJ...846...99M} {846, 99}

\bibitem[\protect\citeauthoryear{{Moschou}, {Drake}, {Cohen},
  {Alvarado-G{\'o}mez}, {Garraffo}  \& {Fraschetti}}{{Moschou}
  et~al.}{2019}]{moschou2019}
{Moschou} S.-P.,  {Drake} J.~J.,  {Cohen} O.,  {Alvarado-G{\'o}mez} J.~D.,
  {Garraffo} C.,   {Fraschetti} F.,  2019, \mn@doi [ApJ]
  {10.3847/1538-4357/ab1b37}, \href
  {https://ui.adsabs.harvard.edu/abs/2019ApJ...877..105M} {877, 105}

\bibitem[\protect\citeauthoryear{{Mu{\~n}oz-Jaramillo}, {Sheeley}, {Zhang}  \&
  {DeLuca}}{{Mu{\~n}oz-Jaramillo} et~al.}{2012}]{munozjaramillo2012}
{Mu{\~n}oz-Jaramillo} A.,  {Sheeley} N.~R.,  {Zhang} J.,   {DeLuca} E.~E.,
  2012, \mn@doi [ApJ] {10.1088/0004-637X/753/2/146}, \href
  {https://ui.adsabs.harvard.edu/abs/2012ApJ...753..146M} {753, 146}

\bibitem[\protect\citeauthoryear{{Muheki}, {Guenther}, {Mutabazi}  \&
  {Jurua}}{{Muheki} et~al.}{2020}]{muheki2020}
{Muheki} P.,  {Guenther} E.~W.,  {Mutabazi} T.,   {Jurua} E.,  2020, \mn@doi
  [MNRAS] {10.1093/mnras/staa3152}, \href
  {https://ui.adsabs.harvard.edu/abs/2020MNRAS.499.5047M} {499, 5047}

\bibitem[\protect\citeauthoryear{{Myers}, {Yamada}, {Ji}, {Yoo}, {Fox},
  {Jara-Almonte}, {Savcheva}  \& {Deluca}}{{Myers} et~al.}{2015}]{myers2015}
{Myers} C.~E.,  {Yamada} M.,  {Ji} H.,  {Yoo} J.,  {Fox} W.,  {Jara-Almonte}
  J.,  {Savcheva} A.,   {Deluca} E.~E.,  2015, \mn@doi [Nature]
  {10.1038/nature16188}, \href
  {https://ui.adsabs.harvard.edu/abs/2015Natur.528..526M} {528, 526}

\bibitem[\protect\citeauthoryear{{Myers}, {Yamada}, {Ji}, {Yoo}, {Jara-Almonte}
   \& {Fox}}{{Myers} et~al.}{2016}]{myers2016}
{Myers} C.~E.,  {Yamada} M.,  {Ji} H.,  {Yoo} J.,  {Jara-Almonte} J.,   {Fox}
  W.,  2016, \mn@doi [Phys. Plasmas] {10.1063/1.4966691}, \href
  {https://ui.adsabs.harvard.edu/abs/2016PhPl...23k2102M} {23, 112102}

\bibitem[\protect\citeauthoryear{{Odert}, {Leitzinger}, {Hanslmeier}  \&
  {Lammer}}{{Odert} et~al.}{2017}]{odert2017}
{Odert} P.,  {Leitzinger} M.,  {Hanslmeier} A.,   {Lammer} H.,  2017, \mn@doi
  [MNRAS] {10.1093/mnras/stx1969}, \href
  {https://ui.adsabs.harvard.edu/abs/2017MNRAS.472..876O} {472, 876}

\bibitem[\protect\citeauthoryear{{Olmedo} \& {Zhang}}{{Olmedo} \&
  {Zhang}}{2010}]{olmedo2010}
{Olmedo} O.,  {Zhang} J.,  2010, \mn@doi [ApJ] {10.1088/0004-637X/718/1/433},
  \href {https://ui.adsabs.harvard.edu/abs/2010ApJ...718..433O} {718, 433}

\bibitem[\protect\citeauthoryear{{Osten} \& {Wolk}}{{Osten} \&
  {Wolk}}{2017}]{osten2017}
{Osten} R.~A.,  {Wolk} S.~J.,  2017, in {Nandy} D.,  {Valio} A.,   {Petit} P.,
  eds,  IAU Symposium Vol. 328, Living Around Active Stars. pp 243--251,
  \mn@doi{10.1017/S1743921317004252}

\bibitem[\protect\citeauthoryear{{Panja}, {Cameron}  \& {Solanki}}{{Panja}
  et~al.}{2020}]{panja2020}
{Panja} M.,  {Cameron} R.,   {Solanki} S.~K.,  2020, \mn@doi [ApJ]
  {10.3847/1538-4357/ab8230}, \href
  {https://ui.adsabs.harvard.edu/abs/2020ApJ...893..113P} {893, 113}

\bibitem[\protect\citeauthoryear{{Patsourakos}, {Vourlidas}  \&
  {Stenborg}}{{Patsourakos} et~al.}{2013}]{patsourakos2013}
{Patsourakos} S.,  {Vourlidas} A.,   {Stenborg} G.,  2013, \mn@doi [ApJ]
  {10.1088/0004-637X/764/2/125}, \href
  {https://ui.adsabs.harvard.edu/abs/2013ApJ...764..125P} {764, 125}

\bibitem[\protect\citeauthoryear{{Patsourakos} et~al.,}{{Patsourakos}
  et~al.}{2020}]{patsourakos2020}
{Patsourakos} S.,  et~al., 2020, \mn@doi [Space Sci. Rev.]
  {10.1007/s11214-020-00757-9}, \href
  {https://ui.adsabs.harvard.edu/abs/2020SSRv..216..131P} {216, 131}

\bibitem[\protect\citeauthoryear{{Petit}, {Louge}, {Th{\'e}ado}, {Paletou},
  {Manset}, {Morin}, {Marsden}  \& {Jeffers}}{{Petit} et~al.}{2014}]{petit2014}
{Petit} P.,  {Louge} T.,  {Th{\'e}ado} S.,  {Paletou} F.,  {Manset} N.,
  {Morin} J.,  {Marsden} S.~C.,   {Jeffers} S.~V.,  2014, \mn@doi [PASP]
  {10.1086/676976}, \href
  {https://ui.adsabs.harvard.edu/abs/2014PASP..126..469P} {126, 469}

\bibitem[\protect\citeauthoryear{{Piskunov} \& {Kochukhov}}{{Piskunov} \&
  {Kochukhov}}{2002}]{piskunov2002}
{Piskunov} N.,  {Kochukhov} O.,  2002, \mn@doi [A\&A]
  {10.1051/0004-6361:20011517}, \href
  {https://ui.adsabs.harvard.edu/abs/2002A&A...381..736P} {381, 736}

\bibitem[\protect\citeauthoryear{{Reiners}}{{Reiners}}{2012}]{reiners2012}
{Reiners} A.,  2012, \mn@doi [Living Rev. Sol. Phys.] {10.12942/Living Rev.
  Solar Phys.-2012-1}, \href {https://ui.adsabs.harvard.edu/abs/2012Living Rev.
  Solar Phys.....9....1R} {9, 1}

\bibitem[\protect\citeauthoryear{{R{\'e}ville}, {Brun}, {Strugarek}, {Matt},
  {Bouvier}, {Folsom}  \& {Petit}}{{R{\'e}ville} et~al.}{2015}]{reville2015}
{R{\'e}ville} V.,  {Brun} A.~S.,  {Strugarek} A.,  {Matt} S.~P.,  {Bouvier} J.,
   {Folsom} C.~P.,   {Petit} P.,  2015, \mn@doi [ApJ]
  {10.1088/0004-637X/814/2/99}, \href
  {https://ui.adsabs.harvard.edu/abs/2015ApJ...814...99R} {814, 99}

\bibitem[\protect\citeauthoryear{{R{\'e}ville}, {Folsom}, {Strugarek}  \&
  {Brun}}{{R{\'e}ville} et~al.}{2016}]{reville2016}
{R{\'e}ville} V.,  {Folsom} C.~P.,  {Strugarek} A.,   {Brun} A.~S.,  2016,
  \mn@doi [ApJ] {10.3847/0004-637X/832/2/145}, \href
  {https://ui.adsabs.harvard.edu/abs/2016ApJ...832..145R} {832, 145}

\bibitem[\protect\citeauthoryear{{Sarkar} \& {Srivastava}}{{Sarkar} \&
  {Srivastava}}{2018}]{sarkar2018}
{Sarkar} R.,  {Srivastava} N.,  2018, \mn@doi [Sol. Phys.]
  {10.1007/s11207-017-1235-8}, \href
  {https://ui.adsabs.harvard.edu/abs/2018Sol. Phys...293...16S} {293, 16}

\bibitem[\protect\citeauthoryear{{Schaefer}, {King}  \&
  {Deliyannis}}{{Schaefer} et~al.}{2000}]{schaefer2000}
{Schaefer} B.~E.,  {King} J.~R.,   {Deliyannis} C.~P.,  2000, \mn@doi [ApJ]
  {10.1086/308325}, \href
  {https://ui.adsabs.harvard.edu/abs/2000ApJ...529.1026S} {529, 1026}

\bibitem[\protect\citeauthoryear{{Schatten}, {Wilcox}  \& {Ness}}{{Schatten}
  et~al.}{1969}]{schatten1969}
{Schatten} K.~H.,  {Wilcox} J.~M.,   {Ness} N.~F.,  1969, \mn@doi [Sol. Phys.]
  {10.1007/BF00146478}, \href {https://ui.adsabs.harvard.edu/abs/1969Sol.
  Phys.....6..442S} {6, 442}

\bibitem[\protect\citeauthoryear{{Schrijver}}{{Schrijver}}{2020}]{schrijver2020}
{Schrijver} C.~J.,  2020, \mn@doi [ApJ] {10.3847/1538-4357/ab67c1}, \href
  {https://ui.adsabs.harvard.edu/abs/2020ApJ...890..121S} {890, 121}

\bibitem[\protect\citeauthoryear{{Schrijver} \& {Title}}{{Schrijver} \&
  {Title}}{2001}]{schrijver2001}
{Schrijver} C.~J.,  {Title} A.~M.,  2001, \mn@doi [ApJ] {10.1086/320237}, \href
  {https://ui.adsabs.harvard.edu/abs/2001ApJ...551.1099S} {551, 1099}

\bibitem[\protect\citeauthoryear{{Schrijver}, {De Rosa}  \&
  {Title}}{{Schrijver} et~al.}{2003}]{schrijver2003}
{Schrijver} C.~J.,  {De Rosa} M.~L.,   {Title} A.~M.,  2003, \mn@doi [ApJ]
  {10.1086/374982}, \href
  {https://ui.adsabs.harvard.edu/abs/2003ApJ...590..493S} {590, 493}

\bibitem[\protect\citeauthoryear{{Schuessler} \& {Solanki}}{{Schuessler} \&
  {Solanki}}{1992}]{schuessler1992}
{Schuessler} M.,  {Solanki} S.~K.,  1992, A\&A, \href
  {https://ui.adsabs.harvard.edu/abs/1992A&A...264L..13S} {264, L13}

\bibitem[\protect\citeauthoryear{{See} et~al.,}{{See} et~al.}{2017}]{see2017}
{See} V.,  et~al., 2017, \mn@doi [MNRAS] {10.1093/mnras/stw3094}, \href
  {https://ui.adsabs.harvard.edu/abs/2017MNRAS.466.1542S} {466, 1542}

\bibitem[\protect\citeauthoryear{{See} et~al.,}{{See} et~al.}{2018}]{see2018}
{See} V.,  et~al., 2018, \mn@doi [MNRAS] {10.1093/mnras/stx2599}, \href
  {https://ui.adsabs.harvard.edu/abs/2018MNRAS.474..536S} {474, 536}

\bibitem[\protect\citeauthoryear{{See} et~al.,}{{See} et~al.}{2019}]{see2019}
{See} V.,  et~al., 2019, \mn@doi [ApJ] {10.3847/1538-4357/ab46b2}, \href
  {https://ui.adsabs.harvard.edu/abs/2019ApJ...886..120S} {886, 120}

\bibitem[\protect\citeauthoryear{{Semel}}{{Semel}}{1989}]{semel1989}
{Semel} M.,  1989, A\&A, \href
  {https://ui.adsabs.harvard.edu/abs/1989A&A...225..456S} {225, 456}

\bibitem[\protect\citeauthoryear{{Shibata} et~al.,}{{Shibata}
  et~al.}{2013}]{shibata2013}
{Shibata} K.,  et~al., 2013, \mn@doi [PASJ] {10.1093/pasj/65.3.49}, \href
  {https://ui.adsabs.harvard.edu/abs/2013PASJ...65...49S} {65, 49}

\bibitem[\protect\citeauthoryear{{Solanki}, {Inhester}  \&
  {Sch{\"u}ssler}}{{Solanki} et~al.}{2006}]{solanki2006}
{Solanki} S.~K.,  {Inhester} B.,   {Sch{\"u}ssler} M.,  2006, \mn@doi [Rep.
  Progr. Phys.] {10.1088/0034-4885/69/3/R02}, \href
  {https://ui.adsabs.harvard.edu/abs/2006RPPh...69..563S} {69, 563}

\bibitem[\protect\citeauthoryear{{Song}, {Chen}, {Ye}, {Han}, {Du}, {Li},
  {Zhang}  \& {Hu}}{{Song} et~al.}{2013}]{song2013}
{Song} H.~Q.,  {Chen} Y.,  {Ye} D.~D.,  {Han} G.~Q.,  {Du} G.~H.,  {Li} G.,
  {Zhang} J.,   {Hu} Q.,  2013, \mn@doi [ApJ] {10.1088/0004-637X/773/2/129},
  \href {https://ui.adsabs.harvard.edu/abs/2013ApJ...773..129S} {773, 129}

\bibitem[\protect\citeauthoryear{{Stenflo} \& {Kosovichev}}{{Stenflo} \&
  {Kosovichev}}{2012}]{stenflo2012}
{Stenflo} J.~O.,  {Kosovichev} A.~G.,  2012, \mn@doi [ApJ]
  {10.1088/0004-637X/745/2/129}, \href
  {https://ui.adsabs.harvard.edu/abs/2012ApJ...745..129S} {745, 129}

\bibitem[\protect\citeauthoryear{{Strassmeier}}{{Strassmeier}}{2009}]{strassmeier2009}
{Strassmeier} K.~G.,  2009, \mn@doi [A\&ARv] {10.1007/s00159-009-0020-6}, \href
  {https://ui.adsabs.harvard.edu/abs/2009A&ARv..17..251S} {17, 251}

\bibitem[\protect\citeauthoryear{{Sun} et~al.,}{{Sun} et~al.}{2015}]{sun2015}
{Sun} X.,  et~al., 2015, \mn@doi [ApJL] {10.1088/2041-8205/804/2/L28}, \href
  {https://ui.adsabs.harvard.edu/abs/2015ApJ...804L..28S} {804, L28}

\bibitem[\protect\citeauthoryear{{T{\"o}r{\"o}k} \& {Kliem}}{{T{\"o}r{\"o}k} \&
  {Kliem}}{2003}]{torok2003}
{T{\"o}r{\"o}k} T.,  {Kliem} B.,  2003, \mn@doi [A\&A]
  {10.1051/0004-6361:20030692}, \href
  {https://ui.adsabs.harvard.edu/abs/2003A&A...406.1043T} {406, 1043}

\bibitem[\protect\citeauthoryear{{T{\"o}r{\"o}k} \& {Kliem}}{{T{\"o}r{\"o}k} \&
  {Kliem}}{2005}]{torok2005}
{T{\"o}r{\"o}k} T.,  {Kliem} B.,  2005, \mn@doi [ApJL] {10.1086/462412}, \href
  {https://ui.adsabs.harvard.edu/abs/2005ApJ...630L..97T} {630, L97}

\bibitem[\protect\citeauthoryear{{T{\"o}r{\"o}k} \& {Kliem}}{{T{\"o}r{\"o}k} \&
  {Kliem}}{2007}]{torok2007}
{T{\"o}r{\"o}k} T.,  {Kliem} B.,  2007, \mn@doi [Astron. Nachr.]
  {10.1002/asna.200710795}, \href
  {https://ui.adsabs.harvard.edu/abs/2007AN....328..743T} {328, 743}

\bibitem[\protect\citeauthoryear{{T{\'o}th}, {van der Holst}  \&
  {Huang}}{{T{\'o}th} et~al.}{2011}]{toth2011}
{T{\'o}th} G.,  {van der Holst} B.,   {Huang} Z.,  2011, \mn@doi [ApJ]
  {10.1088/0004-637X/732/2/102}, \href
  {https://ui.adsabs.harvard.edu/abs/2011ApJ...732..102T} {732, 102}

\bibitem[\protect\citeauthoryear{{Vasantharaju}, {Vemareddy}, {Ravindra}  \&
  {Doddamani}}{{Vasantharaju} et~al.}{2018}]{vasantharaju2018}
{Vasantharaju} N.,  {Vemareddy} P.,  {Ravindra} B.,   {Doddamani} V.~H.,  2018,
  \mn@doi [ApJ] {10.3847/1538-4357/aac272}, \href
  {https://ui.adsabs.harvard.edu/abs/2018ApJ...860...58V} {860, 58}

\bibitem[\protect\citeauthoryear{{Veronig}, {Odert}, {Leitzinger}, {Dissauer},
  {Fleck}  \& {Hudson}}{{Veronig} et~al.}{2021}]{veronig2021}
{Veronig} A.~M.,  {Odert} P.,  {Leitzinger} M.,  {Dissauer} K.,  {Fleck} N.~C.,
    {Hudson} H.~S.,  2021, \mn@doi [Nat. Astron.] {10.1038/s41550-021-01345-9},
  \href {https://ui.adsabs.harvard.edu/abs/2021NatAs...5...72V} {5, 697}

\bibitem[\protect\citeauthoryear{{Vida}, {Leitzinger}, {Kriskovics}, {Seli},
  {Odert}, {Kov{\'a}cs}, {Korhonen}  \& {van Driel-Gesztelyi}}{{Vida}
  et~al.}{2019}]{vida2019}
{Vida} K.,  {Leitzinger} M.,  {Kriskovics} L.,  {Seli} B.,  {Odert} P.,
  {Kov{\'a}cs} O.~E.,  {Korhonen} H.,   {van Driel-Gesztelyi} L.,  2019,
  \mn@doi [A\&A] {10.1051/0004-6361/201834264}, \href
  {https://ui.adsabs.harvard.edu/abs/2019A&A...623A..49V} {623, A49}

\bibitem[\protect\citeauthoryear{{Vidotto}, {Jardine}, {Morin}, {Donati},
  {Opher}  \& {Gombosi}}{{Vidotto} et~al.}{2014}]{vidotto2014}
{Vidotto} A.~A.,  {Jardine} M.,  {Morin} J.,  {Donati} J.~F.,  {Opher} M.,
  {Gombosi} T.~I.,  2014, \mn@doi [MNRAS] {10.1093/mnras/stt2265}, \href
  {https://ui.adsabs.harvard.edu/abs/2014MNRAS.438.1162V} {438, 1162}

\bibitem[\protect\citeauthoryear{{Villadsen} \& {Hallinan}}{{Villadsen} \&
  {Hallinan}}{2019}]{villadsen2019}
{Villadsen} J.,  {Hallinan} G.,  2019, \mn@doi [ApJ]
  {10.3847/1538-4357/aaf88e}, \href
  {https://ui.adsabs.harvard.edu/abs/2019ApJ...871..214V} {871, 214}

\bibitem[\protect\citeauthoryear{{Vourlidas}, {Lynch}, {Howard}  \&
  {Li}}{{Vourlidas} et~al.}{2013}]{vourlidas2013}
{Vourlidas} A.,  {Lynch} B.~J.,  {Howard} R.~A.,   {Li} Y.,  2013, \mn@doi
  [Sol. Phys.] {10.1007/s11207-012-0084-8}, \href
  {https://ui.adsabs.harvard.edu/abs/2013SoPh..284..179V} {284, 179}

\bibitem[\protect\citeauthoryear{{Vr{\v{s}}nak}}{{Vr{\v{s}}nak}}{2001}]{vrsnak2001}
{Vr{\v{s}}nak} B.,  2001, \mn@doi [J. Geophys. Res.] {10.1029/2000JA004007},
  \href {https://ui.adsabs.harvard.edu/abs/2001JGR...10625249V} {106, 25249}

\bibitem[\protect\citeauthoryear{{Vr{\v{s}}nak}, {Sudar}  \&
  {Ru{\v{z}}djak}}{{Vr{\v{s}}nak} et~al.}{2005}]{vrsnak2005}
{Vr{\v{s}}nak} B.,  {Sudar} D.,   {Ru{\v{z}}djak} D.,  2005, \mn@doi [A\&A]
  {10.1051/0004-6361:20042166}, \href
  {https://ui.adsabs.harvard.edu/abs/2005A&A...435.1149V} {435, 1149}

\bibitem[\protect\citeauthoryear{{Wang} \& {Sheeley}}{{Wang} \&
  {Sheeley}}{1989}]{wang1989}
{Wang} Y.~M.,  {Sheeley} N.~R. J.,  1989, \mn@doi [Sol. Phys.]
  {10.1007/BF00146521}, \href {https://ui.adsabs.harvard.edu/abs/1989Sol.
  Phys...124...81W} {124, 81}

\bibitem[\protect\citeauthoryear{{Wang} \& {Sheeley}}{{Wang} \&
  {Sheeley}}{1992}]{wang1992}
{Wang} Y.~M.,  {Sheeley} N.~R. J.,  1992, \mn@doi [ApJ] {10.1086/171430}, \href
  {https://ui.adsabs.harvard.edu/abs/1992ApJ...392..310W} {392, 310}

\bibitem[\protect\citeauthoryear{{Wang} \& {Zhang}}{{Wang} \&
  {Zhang}}{2007}]{wangy2007}
{Wang} Y.,  {Zhang} J.,  2007, \mn@doi [ApJ] {10.1086/519765}, \href
  {https://ui.adsabs.harvard.edu/abs/2007ApJ...665.1428W} {665, 1428}

\bibitem[\protect\citeauthoryear{{Wang}, {Liu}, {Wang}, {Liu}, {Chen}, {Liu},
  {Zhou}  \& {Zhang}}{{Wang} et~al.}{2017}]{wangd2017}
{Wang} D.,  {Liu} R.,  {Wang} Y.,  {Liu} K.,  {Chen} J.,  {Liu} J.,  {Zhou} Z.,
    {Zhang} M.,  2017, \mn@doi [ApJL] {10.3847/2041-8213/aa79f0}, \href
  {https://ui.adsabs.harvard.edu/abs/2017ApJ...843L...9W} {843, L9}

\bibitem[\protect\citeauthoryear{{Webb} \& {Howard}}{{Webb} \&
  {Howard}}{2012}]{webb2012}
{Webb} D.~F.,  {Howard} T.~A.,  2012, \mn@doi [Living Rev. Sol. Phys.]
  {10.12942/Living Rev. Solar Phys.-2012-3}, \href
  {https://ui.adsabs.harvard.edu/abs/2012Living Rev. Solar Phys.....9....3W}
  {9, 3}

\bibitem[\protect\citeauthoryear{{Wood} et~al.,}{{Wood}
  et~al.}{2021}]{wood2021}
{Wood} B.~E.,  et~al., 2021, \mn@doi [ApJ] {10.3847/1538-4357/abfda5}, \href
  {https://ui.adsabs.harvard.edu/abs/2021ApJ...915...37W} {915, 37}

\bibitem[\protect\citeauthoryear{{Yashiro}, {Akiyama}, {Gopalswamy}  \&
  {Howard}}{{Yashiro} et~al.}{2006}]{yashiro2006}
{Yashiro} S.,  {Akiyama} S.,  {Gopalswamy} N.,   {Howard} R.~A.,  2006, \mn@doi
  [ApJL] {10.1086/508876}, \href
  {https://ui.adsabs.harvard.edu/abs/2006ApJ...650L.143Y} {650, L143}

\bibitem[\protect\citeauthoryear{{Yeates}}{{Yeates}}{2020}]{yeates2020}
{Yeates} A.~R.,  2020, \mn@doi [Sol. Phys.] {10.1007/s11207-020-01688-y}, \href
  {https://ui.adsabs.harvard.edu/abs/2020Sol. Phys...295..119Y} {295, 119}

\bibitem[\protect\citeauthoryear{{Zhang}, {Dere}, {Howard}, {Kundu}  \&
  {White}}{{Zhang} et~al.}{2001}]{zhang2001}
{Zhang} J.,  {Dere} K.~P.,  {Howard} R.~A.,  {Kundu} M.~R.,   {White} S.~M.,
  2001, \mn@doi [ApJ] {10.1086/322405}, \href
  {https://ui.adsabs.harvard.edu/abs/2001ApJ...559..452Z} {559, 452}

\bibitem[\protect\citeauthoryear{{Zhou}, {Cheng}, {Zhang}, {Wang}, {Wang},
  {Liu}, {Zhuang}  \& {Cui}}{{Zhou} et~al.}{2019}]{zhouzj2019}
{Zhou} Z.,  {Cheng} X.,  {Zhang} J.,  {Wang} Y.,  {Wang} D.,  {Liu} L.,
  {Zhuang} B.,   {Cui} J.,  2019, \mn@doi [ApJL] {10.3847/2041-8213/ab21cb},
  \href {https://ui.adsabs.harvard.edu/abs/2019ApJ...877L..28Z} {877, L28}

\bibitem[\protect\citeauthoryear{{Zic} et~al.,}{{Zic} et~al.}{2020}]{zic2020}
{Zic} A.,  et~al., 2020, \mn@doi [ApJ] {10.3847/1538-4357/abca90}, \href
  {https://ui.adsabs.harvard.edu/abs/2020ApJ...905...23Z} {905, 23}

\bibitem[\protect\citeauthoryear{{Zuccarello}, {Aulanier}  \&
  {Gilchrist}}{{Zuccarello} et~al.}{2015}]{zuccarello2015}
{Zuccarello} F.~P.,  {Aulanier} G.,   {Gilchrist} S.~A.,  2015, \mn@doi [ApJ]
  {10.1088/0004-637X/814/2/126}, \href
  {https://ui.adsabs.harvard.edu/abs/2015ApJ...814..126Z} {814, 126}

\bibitem[\protect\citeauthoryear{{van Ballegooijen}, {Cartledge}  \&
  {Priest}}{{van Ballegooijen} et~al.}{1998}]{vanballegooijen1998}
{van Ballegooijen} A.~A.,  {Cartledge} N.~P.,   {Priest} E.~R.,  1998, \mn@doi
  [ApJ] {10.1086/305823}, \href
  {https://ui.adsabs.harvard.edu/abs/1998ApJ...501..866V} {501, 866}

\makeatother
\end{thebibliography}








\bsp	
\label{lastpage}
\end{document}